\documentclass[%
 reprint,
 amsmath,amssymb,
aps,
prstab,
]{revtex4-1}
\usepackage{caption}
\usepackage{placeins}
\usepackage{graphicx}
\usepackage{dcolumn}
\usepackage{bm}
\begin{document}
\captionsetup[figure]{justification=raggedright}
\captionsetup[table]{justification=raggedright}
\preprint{APS/123-QED}

\title{Theoretical and Simulation Study of $'$Comb$'$ Electron beam and
	THz generation}

\author{V. Joshi}\email{Electronic Address: vipuljoshi92@gmail.com}\author{J. Karmakar} \author{N. Kumar}\author{ B. Karmakar} \author{S. Tripathi}\author{S. Ghosh} \author{R.K. Bhandari}\author{ D. Kanjilal}%
 \affiliation{%
 	Inter University Accelerator Centre (IUAC), Aruna Asaf Ali Marg, New Delhi, India 110067 
 }%

\author{U. Lehnert}
\affiliation{
 Helmholtz Zentrum Rossendorf Dresden (HZDR), Dresden, Germany
}%

\author{A. Aryshev, J. Urakawa}
\affiliation{%
 High Energy Accelerator Research Organization, KEK, Tsukuba, Japan
}%

\date{\today}

\begin{abstract}
A compact accelerator based super-radiant THz source is under development at Inter University Accelerator Centre (IUAC), New Delhi. The facility is based on the principle of pre-bunched Free Electron Laser (FEL) which will produce THz radiation in the range of $\sim$0.18 to 3 THz from a modulated electron beam. A photocathode electron gun will generate a short train of micro-bunches (a $"$comb$"$ beam) driven by a fiber laser system capable of producing multi micro-pulse laser beam with variable separation ( $"$comb$"$ laser pulse). Upon acceleration, the electron beam will be injected in to a compact undulator magnet tuned to the same frequency as the separation of the electron micro-bunches.
The paper discusses the process of enhancement of super-radiant emission of radiation due to modulation in the comb beam and the conditions required to achieve maximum enhancement of the radiation power. The feasibility study of generating a comb beam at the photocathode and its transport through the beamline while preserving its temporal structure has been reported. To evaluate the characteristics of the radiation emitted from the comb beam, a $C^+$$^+$ based particle tracker and Li\~enard-Wiechert field solver has been developed. The conceptual understanding of the emission of radiation from comb beam is shown to conform with the numerical results. The code has been used to calculate the radiation pulse energy emitted into the central cone of undulator for various comb beam configurations.

\end{abstract}
\maketitle
\section{\label{sec:level1}Introduction}
The advent of the THz sources and the advantages of THz time-domain spectroscopy over the Far Infra-Red (FIR) Fourier transform spectroscopy has placed the scientific world "At the Dawn of the New Era in THz Technology" \cite{Hosako2007}. Until recently, the availability of THz sources was limited and the region between 0.1 THz to 10 THz was called as 'THz Gap'. However, with the advancements in the schemes of producing THz sources and its indispensable futuristic roles in understanding plethora of physical, chemical and biological processes, the THz Gap is firming its position as bridge between the fundamental sciences e.g. non-linear optics \cite{Kono1997,Rojan2017,Danielson2007}, condensed matter physics \cite{Zalden2016,Ulbricht2011}, non-linear spectroscopy \cite{Jewariya2010}, selective control of magnetic properties of materials \cite{Kovalev2018} etc. and the real world applications like characterization of materials \cite{Ferguson2002},  non-invasive imaging \cite{Knap2009}, national security \cite{Shen2005,Zandonella2003,Stoik2008}, THz communication \cite{Koenig2013}, monitoring of industrial processes \cite{Skryl2014,Duling2009}, biomedical applications \cite{Yu2012} etc.    \par 
The required boost to the development of THz sources is attributed to immense research activities in the fields of Photo-conductive switches \cite{Auston1975,Auston1984,You1993,Budiarto1996}, Optical Rectification \cite{Bass1962,Yang1971,Rice1994}, Accelerator based Coherent Synchrotron Radiation (CSR) \cite{Michel1982,Nakazato1989,Bocek1996,Zhang2012,Perucchi2013,Green2016,Nasse2013,Gensch2008} or Coherent Transition Radiation (CTR) \cite{Casalbuoni2009,Hoffmann2011a,Wu2013,Carr2002}, Quantum Cascade Lasers \cite{Williams2007} and Laser Induced Air Plasma \cite{Xie2006} based devices. Amongst the variety of THz sources, Photoconductive switches and Optical Rectification techniques are most commonly used and historically important. Photoconductive switches are based on biased semiconductor devices (Auston Switches \cite{Auston1975,Auston1984}  ) which can be thought of as loaded capacitor \cite{Hoffmann2011}. If a femtosecond laser pulse is incident on the switch, the  stored electrostatic energy is released in the form of ultrashort, high bandwidth, single-cycle THz radiation having pulse energies in the range of few femtojoule per pulse. Much higher THz pulse energies ($\sim$ 0.8 $\mu$J) using photo-conductive switches were obtained by \textit{You et al.} by applying high-bias voltage across GaAs crystalline wafers and exciting the switch by 120 fs, 770 nm laser pulses from a Ti:Sapphire chirped-pulse amplifier system \cite{You1993}. The most common mechanism to generate single cycle (broadband) or multi-cycle (narrowband) THz pulses having $\sim \mu$J pulse energies is to use the Optical Rectification (OR) technique \cite{Yang1971} which is based on inducing instantaneous second-order nonlinear susceptibility $\chi^2$ (and therefore non-linear time-dependent polarization $P^2(t)$) in an organic or inorganic semiconductor crystal by using an intense broadband femtosecond pump pulse. Due to second-order non-linear processes inside the crystal, intrapulse difference frequency generation (DFG) occurs; resulting in a rectified THz pulse \cite{Hoffmann2011,Lee2000}. The most commonly used crystal for OR is ZnTe while other crystals e.g. GaP, GaAs, DAST, LiNb$O_3$ are becoming increasingly popular. Recently, several groups have shown that OR techniques can be used to generate THz pulses with high fields ($>$1 MV/cm) \cite{Hirori2011,Oh2014,Hebling2008}, high pulse energies \cite{Yeh2007} and multi-cycle narrowband THz pulses \cite{Lee2000}.   \par
 Although THz sources based on Photoconductive switches or OR are more accessible, less bulky and commercially available; accelerator based THz sources offers much higher brilliance, pulse energies, tunability, repetition rates and offer both broadband (from bending magnets) and narrowband THz pulses (from undulators) \cite{Stojanovic2013}. The possibility of intense coherent emission of radiation from short bunches of relativistic particles in a storage ring was first mentioned by Michel in 1982 \cite{Michel1982} and observed by  Nakazato at Tohoku in 1989 \cite{Nakazato1989}. Over the last two decades, there has been significant improvement in the techniques employed to generate high power narrowband tunable coherent THz pulses from storage rings. \textit{K. Holldack et. al} have used energy modulated electron beam ('femtosliced' electron beam) to produce femtosecond broadband THz pulses at BESSY storage ring \cite{Holldack2006}.   \textit{J. M. Byrd et al.} \cite{Byrd2006} at Advanced Light Source has been able to produce temporally and spatially coherent THz pulses using longitudinally and transversally modulated electron beam. They also showed that the electric field of the THz pulses can be tailored by varying the intensity, delay and number of pulses of the slicing laser pulse . Groups at UVSOR at Okazaki, Japan and DELTA in Dortmund, Germany have demonstrated generation of continuously tunable narrowband pulses in the THz and sub-THz regime using laser modulated electron beams \cite{Bielawski2008,Ungelenk2017}. \par
 In 1979, John M.J. Madey invented FEL \cite{Madey1971} and his group successfully demonstrated its operation in 1977  at Stanford \cite{Deacon1977}. The world's first FEL working in the THz regime was based on electrostatic accelerator at the University of California, Santa Barbara (UCSB) FEL . The original FEL at UCSB is now replaced with two FELs (MM-FEL and FIR-FEL) and cover the frequency range of 0.12 THz to 4.8 THz \cite{Ramian1992}. Another electrostatic accelerator based FEL is the Israeli FEL which can operate in continuous wave (cw) or quasi-static mode. The Israeli FEL achieved its first lasing in 1997 at pulsed power mode and provided 1kW of power at 0.1 THz \cite{Abramovich1997}. The Free Electron Lasers at Jefferson Lab, Virginia (JFEL) \cite{Neil2000} and Novosibirsk, Russia (NovoFEL) \cite{Knyazev2010} are based on Energy Recovery Linacs (ERL) and have been able to provide infrared (IR) or THz pulses at 1.72 kW and 500 W respectively.  Except for ERL based FELs or Electrostatic Accelerator based FELs, cw mode operation of FELs can be achieved in super-conducting linacs only. The \textbf{F}EL at the \textbf{E}lectron\textbf{ L}inear accelerator with high \textbf{B}rilliance and Low \textbf{E}mittance (FELBE) facility at Helmholtz-Zentrum Dresden-Rossendorf (HZDR) Dresden, Germany has a thermionic gun coupled with two super-conducting linac modules which operates in a cw or quasi cw mode with 15 MHz pulse repetition rate. Using two FELs (U37-FEL and U100-FEL), they are able to cover mid and far infrared regime of electromagnetic spectrum  \cite{Winnerl2006}. Also, TELBE facility at HZDR can produce multi-cycle THz pulses at very high repetition rates. TELBE facility have a thermionic gun based electron injector which can be operated at 13 MHz repetition rate with 100 pC bunch charge to produce THz pulse energy of the order of 1 $\mu$J and is working on SRF injector based THz source which shall operate at 500 kHz repetition rate with 1 nC bunch charge to produce pulse energies as high as 100 $\mu$J\cite{Green2016}. For a more detailed list of FELs, readers are referred to a review article titled 'Accelerator Sources for THz Science: A Review' by G.R. Neil \cite{Neil2014}.\par
 The latest entry to the accelerator based THz sources are the Linac based Super-Radiant THz sources. It is now very well understood that the quality of the emitted radiation from a collection of electrons depends upon the number of electrons in the bunch and the phase-space volume occupied by the  electrons in the bunch \cite{Gover2005}. Intense coherent radiation can be generated from electron bunches if (a) the electron beam bunch length is much smaller than the radiation wavelength ($\lambda$) to be generated, (b) the transverse beam size is of the order of $\lambda/4\pi$ and (c) the number of the particles in the bunch is as high as possible. Under these conditions, the radiation power scales as the square of the number of the particles in the bunch \cite{Hirschmugl1991}. Several groups have used super-radiant electron bunches to produce intense  CTR \cite{Wu2013,Kung1994,Hoffmann2011a} or Coherent Undulator Radiation \cite{Green2016} in the THz regime.    \par
 In this paper, a compact THz source project at IUAC, named as Delhi Light Source (DLS) \cite{Ghosh2017}, capable of producing continuously tunable, narrowband, multi-cycle, coherent THz pulses in the range of $\sim$0.18 to 3 THz from a train of super-radiant micro-bunches (henceforth called as $"$comb$"$ beam) is being presented. In the present approach, we show that train of super-radiant micro-bunches can be produced from a normal conducting photocathode electron gun by temporally modulating the incident photolaser; eliminating the need of a magnetic chicane to compress the electron bunch and reducing the size of the beamline. Further, the flexibility of varying the separation between the micro-bunches, the total number of micro-bunches and the charge per micro-bunch of the comb beam gives an additional advantage to provide tunable frequencies, variable intensities and variable pulse duration of the electromagnetic radiation. \par 
 \begin{figure*}
 	\fbox{\includegraphics[width=15.92cm,keepaspectratio]{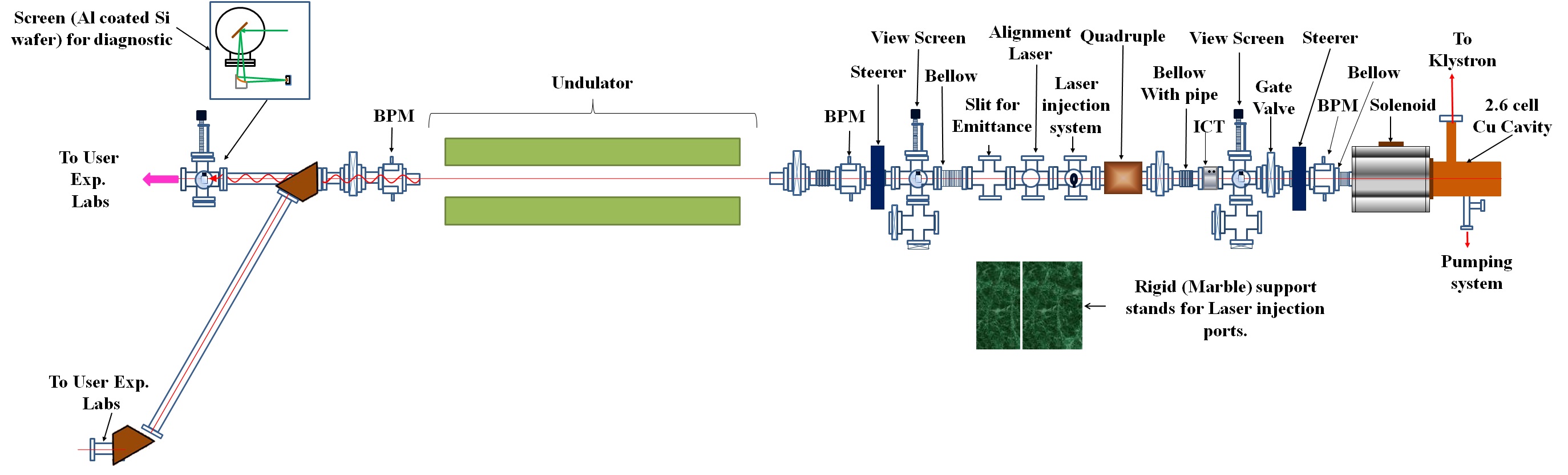}}
 	\captionof{figure}{\label{fig:dls_beamline}Layout of beamline for Phase-1 of Delhi Light Source.}
 \end{figure*}
 Presently, effort is on to develop the first phase of DLS which is briefly described in Section \ref{Sec:DLS}. In Section \ref{Theory}, the conditions required to enhance the radiation power emitted into the resonant frequency of the undulator by the comb electron beam has been discussed. The results of beam optics calculations done by using the General Particle Tracer (GPT) \cite{Geer1996} code has been presented in Section \ref{BeamOptics}. To simulate the emission of the radiation inside the Undulator, a code based on $C^+$$^+$ has been developed to track the electron beam inside the Undulator and to compute the radiation emitted by them using the Li\~enard-Wiechert fields \cite{Jefimenko1992,Jackson1999}. The results of the THz calculations has been presented in Section \ref{THz calculations}.
 \section{\label{Sec:DLS} Delhi Light Source: Phase-1} 
  The first phase of Delhi Light Source, is an ongoing project at IUAC to develop a THz radiation source facility. The complete facility will be developed in three phases and the work towards the Phase 1 (layout shown in \ref{fig:dls_beamline}) is presently going on. The aim of the first phase is to develop a compact super-radiant THz source based on normal conducting (OFHC copper) photo-injector for pulsed mode operation \cite{Ghosh2017}.   \par
\begin{figure}
	\fbox{\includegraphics[width=8.75cm,keepaspectratio]{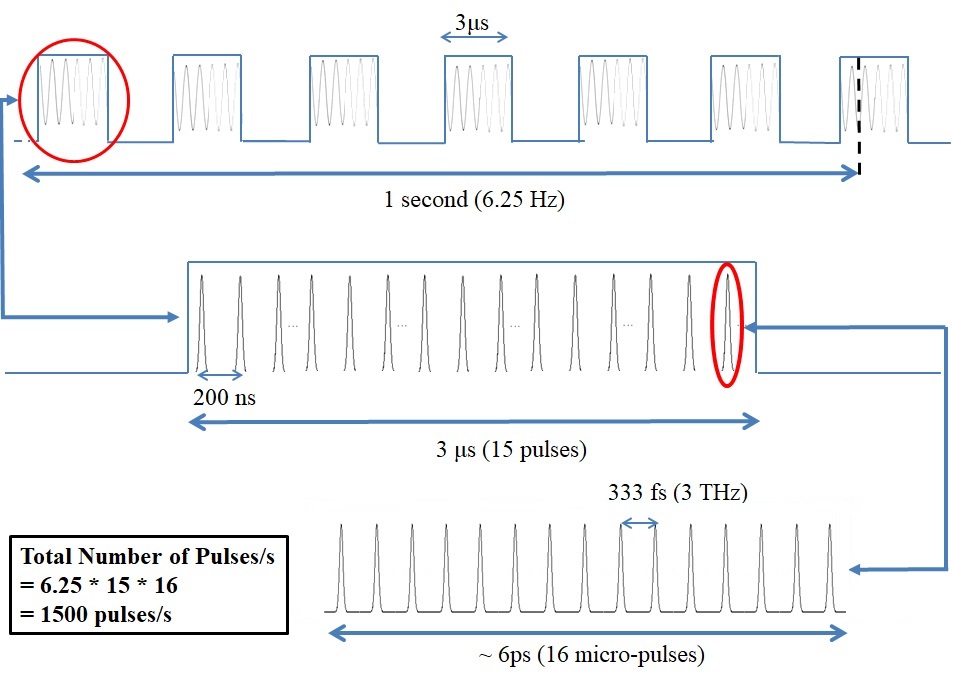}}
	\captionof{figure}{\label{CarrierEnvelope} Carrier Envelope of the multi micro-pulse structure for low duty cycle (Phase-1) operation of photoinjector. The repetition rate of the macropulses is 6.25 Hz (top) having 15 pulses in each macropulse (middle) and all of the 15 pulses are split into multi micro-pulses (bottom). }
\end{figure}

 
  The choice of the Laser system for DLS is a Yb
  doped fiber laser system with an oscillator frequency of 130MHz ($10^{th}$ sub-harmonic of the
  main master clock) and 1030 nm as fundamental wavelength. However, 130 MHz repetition
  rate will be reduced to 5 MHz to reduce laser pulse loading inside the fiber. At this point it is
  important to mention that the normal conducting cavity in the Phase-1 of the DLS project will
  be operated in a low duty cycle mode and therefore the length of the RF window is limited to
  $\sim$ 3 $\mu$seconds with 6.25 Hz repetition rate. Thus, in a time window of $\sim$ 3 $\mu$seconds, 15 laser
  pulses will be selected. Each of these 15 laser pulses will be passed through two stage burst
  amplification to increase the laser pulse energy. After this stage, Michelson Interferometer
  type pulse-divider( based on multi-stage polarised beam splitting with optical delay
  lines) \cite{Aryshev2017,Shevelev2017} will split each of the 15 laser pulses into 2, 4, 8 or 16 micro-pulses. The laser micro-pulses will be then converted from fundamental IR (1030nm) to UV
  (258nm) before being incident on the photo-cathode to generate the comb electron beam. The carrier envelope of the electron beam is shown in Fig. \ref{CarrierEnvelope} and the expected pulse energy in a single micro-pulse is given in the Table \ref{tab:PulseEnergies} .   The laser system is being developed in collaboration with KEK, Japan. \par
\begin{table}[h]
  	\caption{\label{tab:PulseEnergies}%
  		The expected pulse energy in a single laser micro-pulse and the maximum amount of charge per micro-bunch that can be produced using $Cs_2Te$ photocathode. 
  	}
  	\begin{ruledtabular}
  		\begin{tabular}{ccccc}
  			\multicolumn{1}{p{1cm}}{\centering Energy \\ $\left(\mu J\right)$}&
  			\multicolumn{1}{p{1cm}}{\centering Number}&
  			\multicolumn{1}{p{1cm}}{\centering FWHM \\(fs)}&
  			\multicolumn{1}{p{1cm}}{\centering Cathode}&
  			\multicolumn{1}{p{2.5cm}}{\centering Max. Charge /micro-bunch \\(pC)}\\
  			\colrule
  			\multicolumn{1}{p{1cm}}{\centering 0.1 \\ (Steady State)}&	1-16&	200&	$Cs_2Te$&	200 
  			
  		\end{tabular}
  	\end{ruledtabular}
  \end{table}
  The accelerating structure for the Phase-1 of DLS is a 2.6 cell, S-band (2860 MHz) normal  conducting resonant cavity which will be powered by a 25 MW high power RF system consisting of Klystron and Modulator. The cavity is coupled with a solenoid to compensate for the growth in emittance. The temporal structure of the comb electron beam will be identical to the comb laser pulses at the cathode and therefore; the accelerating field strength at the cathode surface must be high enough to preserve the modulation of the electron beam through the initial stages of the acceleration until reaching relativistic velocities. The maximum energy of the comb beam is expected to be $\sim$8 MeV.\par 
  The comb beam will be injected into a planar undulator to produce coherent THz radiation. Inside the undulator, the wave packets emitted from individual micro-bunches inside the same micro-bunch train will interfere constructively if the separation between the micro-bunches is made equal to the wavelength of the radiation. If the coherence condition is satisfied then the intensity of the emitted Coherent Undulator Radiation \cite{Neuman2000,Bocek1996} will be proportional to the square of the total number of electrons in the multi micro-bunch structure of the electron beam \cite{Hirschmugl1991}. By adjusting the electron energy and changing the gap of the undulator magnet, the wavelength of the radiation frequency $\lambda_r$ can be adjusted in the range of 0.18 to 3.0 THz by following the equation \cite{Madey1971,Motz1951}:
\begin{equation}
\label{UnduEq}
\lambda_r = \frac{\lambda_u}{2\gamma^2}(1+K^2/2+\gamma^2\theta^2)
\end{equation}
where $K$=0.934 $ B_u[T]\lambda_u[cm]$ is the undulator parameter, $\lambda_u$ is the undulator period, $B_u$  is the peak magnetic field on the axis of the undulator, $\gamma$ is the relativistic factor for the electron and $\theta$ is the angle of observation of the radiation. \par

\section{Theory of Coherent Emission of Radiation from the Comb Beam} \label{Theory}
It is well understood that an accelerating charged particle emits electro-magnetic radiation whose power varies remarkably with the particle$'$s energy. The electric and magnetic fields of the emitted radiation can be found using the Li\~enard-Wiechert formulation for the fields of a point particle of charge q, which is given as \cite{Jefimenko1992,Jackson1999}:
\begin{subequations}\label{LW}
\begin{eqnarray}
		\textbf{E}(\textbf{r},t)= \dfrac{q}{4\pi\epsilon_0}\Bigg(\dfrac{\hat{\textbf{n}}-\boldsymbol{\beta}}{\gamma^2|\textbf{R}|^2 \left(1-\hat{\textbf{n}}\boldsymbol{\cdot}\boldsymbol{\beta}\right)^3} \nonumber\qquad \qquad \qquad \\ + \dfrac{\hat{\textbf{n}}\times \left(\left(\hat{\textbf{n}} -  \boldsymbol{\beta}\right)\times \dot{\boldsymbol{\beta}}\right)}{c|\textbf{R}|\left(1-\hat{\textbf{n}}\boldsymbol{\cdot}\boldsymbol{\beta}\right)^3}\Bigg)_{t_r} 
\end{eqnarray}
\begin{eqnarray}
	\textbf{B}(\textbf{r},t)=\dfrac{\hat{\textbf{n}}\times\textbf{E}(\textbf{r},t)}{c}\Bigg|_{t_r} \qquad\qquad\qquad \qquad\quad\quad\quad
\end{eqnarray}
\end{subequations}
where \textbf{R = r -} \textbf{r}$_s$ is the distance of the observation point (\textbf{r}) from the source of radiation (\textbf{r$_s$}), $\bf{\hat{n}}$ = \textbf{R}/R is the unit vector pointing from source of radiation to the observation point, $\boldsymbol{\beta}$, $\boldsymbol{\dot{\beta}}$ are the velocity and acceleration of the particle respectively, evaluated at retarded time $t_r$.\par
The Li\~enard-Wiechert fields can be expressed in the frequency domain by evaluating the Fourier Transform of Eq. \ref{LW}. Under the approximation that the observation point is extremely far from the source of the radiation (r$_s$ \textless\textless r); we can approximate the unit vector pointing from the electron to observer as $\hat{\textbf{n}}$ $\simeq$$\textbf{ r}$/r. This approximation allows us to write the result of Fourier Transform of the radiation term (second term in Eq.\ref{LW}(a))  as:
\begin{eqnarray}
	\textbf{E}(\textbf{r},\omega)=A(\textbf{r},\omega)\int_{- \infty}^{+\infty} \hat{\textbf{n}} \times \Big(\hat{\textbf{n}}\times \boldsymbol{\beta} \Big) \qquad \qquad \nonumber \quad \\ \exp\Big(i\omega \big(t_r -\hat{\textbf{n}} \boldsymbol{\cdot} \textbf{r}_s/c\big)\Big)dt_r
\end{eqnarray}
where $A(\textbf{r},\omega) = \dfrac{-ik\omega}{r} \exp\Big(\dfrac{i\omega r}{c}\Big)$ and k=$\dfrac{-e}{\sqrt{32\pi^3}\epsilon_0 c}$ is a constant defined for electrons. \par
If instead of a single electron, a bunch of $N_e$ electrons is considered; then the total electric field of the emitted radiation wavepacket is given by the sum of the electric fields emitted by individual electrons:

\begin{eqnarray}
\textbf{E}(\textbf{r},\omega)=A(\textbf{r},\omega)\sum_{j=1}^{N_e}\Bigg(\int_{- \infty}^{+\infty} \hat{\textbf{n}} \times \Big(\hat{\textbf{n}}\times \boldsymbol{\beta_j} \Big) \nonumber \\ \exp\Big(i\omega \big(t_r -\hat{\textbf{n}} \boldsymbol{\cdot} \textbf{r}_{s,j}/c\big)\Big)dt_r\Bigg)
\end{eqnarray} \par
 Following Hirschmugl \cite{Hirschmugl1991}, if we assume that all the electrons move with same velocity i.e. $\beta_j=\beta$, the coordinate of the j$^{th}$ particle with respect to the centre of mass (\textbf{r$_c$}) of the bunch is given by (\textbf{r$_j$}) and the distance of the observer from the centre of mass of the bunch is much larger than the bunch length ( $l_b \textless \textless r$); then the summation in Eq. 4 can be written as:

\begin{eqnarray}
\textbf{E}(\textbf{r},\omega)= A(\textbf{r},\omega) \Bigg(\sum_{j=1}^{N_e} \exp\Big(i\omega\dfrac{\hat{\textbf{n}} \boldsymbol{\cdot} \textbf{r}_j}{c}\Big)\Bigg) \qquad \qquad \nonumber \\ \Bigg(\int_{- \infty}^{+\infty} \hat{\textbf{n}} \times \Big(\hat{\textbf{n}}\times \boldsymbol{\beta} \Big)  \exp\bigg(i\omega \bigg(t_r -\dfrac{\hat{\textbf{n}} \boldsymbol{\cdot} \textbf{r}_c}{c}\bigg)\bigg)dt_r\Bigg)
\end{eqnarray}\par
The experimentally measurable quantity is the intensity of the radiation which depends on the square of the amplitude of the electric field. From Eq. 5, the amplitude of the total radiation wave packet, corresponding to a particular frequency $\omega$,  emitted by a bunch will be:
\begin{eqnarray}
\label{ModE}
\Bigg|\textbf{E}(\textbf{r},\omega)\Bigg|= \Bigg|A(\textbf{r},\omega)\Bigg| \Bigg|\sum_{j=1}^{N_e} \exp\Big(i\omega\dfrac{\hat{\textbf{n}}\boldsymbol{\cdot} \textbf{r}_j}{c}\Big)\Bigg|\qquad \qquad \nonumber \\ \Bigg|\int_{- \infty}^{+\infty} \hat{\textbf{n}} \times \Big(\hat{\textbf{n}}\times \boldsymbol{\beta} \Big)  \exp\bigg(i\omega \bigg(t_r -\dfrac{\hat{\textbf{n}} \boldsymbol{\cdot} \textbf{r}_c}{c}\bigg)\bigg)dt_r\Bigg| \quad
\end{eqnarray}
Now, if we define the bunching factor or form factor \cite{Hirschmugl1991} $B(\omega)$ as 
\begin{equation}
 B(\omega)=\dfrac{\bigg|\displaystyle\sum_{j=1}^{N_e}\exp⁡\Big(i\omega\dfrac{\hat{\textbf{n}}\boldsymbol{\cdot}\textbf{r}_j}{c}\Big)\bigg |}{N_e} 
\end{equation} 
\\ and the integral $I(t_r)$ (the detailed solution of this integral can be found in \cite{G.DattoliA.Renieri1993}) as 
\begin{equation}
I(t_r)=\Bigg|\int_{- \infty}^{+\infty} \hat{\textbf{n}} \times \Big(\hat{\textbf{n}}\times \boldsymbol{\beta} \Big)  \exp\bigg(i\omega \bigg(t_r -\dfrac{\hat{\textbf{n}} \boldsymbol{\cdot} \textbf{r}_c}{c}\bigg)\bigg)dt_r\Bigg|
\end{equation}
\\then Eq. \ref{ModE} becomes:
\begin{eqnarray}
\Big|\textbf{E}(\textbf{r},\omega)\Big|= \Big|A(\textbf{r},\omega)\Big|N_eB(\omega)I(t_r)
\end{eqnarray}
From the above equation, it is clear that the intensity of radiation emitted into a particular mode $\omega$ is directly dependent on the bunching factor (corresponding to the same mode). If the bunch length $l_b$ (which can be approximated as two times the position of the electron farthest from the centre of mass) of the beam is extremely small compared to the wavelength of the emitted radiation ($\omega r_j/c=\pi l_b/\lambda \rightarrow 0)$, then the entire bunch is approximated as a point source of radiation; resulting in maximum possible intensity of radiation. The coherence effects due to this condition was initially termed as $'$super-bunching$'$ \cite{Michel1982} and has since been studied and observed by several authors \cite{Williams1989,Nakazato1989,Deacon1977,Billinghurst2013}.
\par
The above discussion can be further extended to find the electromagnetic fields of  a radiation wavepacket emitted by a comb beam. Let us consider a comb beam in which there are $N_m$ micro-bunches having $N_e$ electrons in each micro-bunch, the position of $j^{th}$ electron (with respect to the centre of mass) in the $m^{th}$ micro-bunch is given by $\bf r_{j,m}$  and that the centre of mass of $m^{th}$ micro-bunch follows the trajectory given by $\bf r_{c,m}$. The electric field of the wavepacket emitted by this beam will be given by:
\begin{eqnarray}
\textbf{E}(\textbf{r},\omega)=A(\textbf{r},\omega) \sum_{m=0}^{N_m-1}\Bigg\{\bigg(\sum_{j=1}^{N_e} \exp\Big(i\omega\dfrac{\hat{\textbf{n}} \boldsymbol{\cdot} \textbf{r}_{j,m}}{c}\Big)\bigg)\nonumber\\
\int_{- \infty}^{+\infty} \hat{\textbf{n}} \times \Big(\hat{\textbf{n}}\times \boldsymbol{\beta_m} \Big)  \exp\bigg(i\omega \bigg(t_r -\dfrac{\hat{\textbf{n}} \boldsymbol{\cdot} \textbf{r}_{c,m}}{c}\bigg)\bigg)dt_r\Bigg\} \nonumber \\
\end{eqnarray}
where $\boldsymbol{\beta_m}$ is the velocity of the $m^{th}$ micr-obunch. 
In the above expression for the electric field, care needs to be taken to separate the integral from the summation. Let the trajectory followed by the centre of mass of the first micro-bunch and its velocity be given by{ $\textbf{r}_c$} and $\boldsymbol{\beta}$ respectively. We assume that the particles of all the micro-bunches are ultra-relativistic and have same velocity i.e. $\boldsymbol{\beta_m$} $\simeq$ $\boldsymbol{\beta}$. The trajectory of centre of mass of all the other micro-bunches can then be defined with respect to the first micro-bunch, if we assume that the relative separation between the micro-bunches ($\boldsymbol{\Delta r_m}$) is fixed and is independent of time. Under this condition, we can write: $\textbf{r}_{c,m}$ = $\textbf{r}_c$ - $\boldsymbol{\Delta}\textbf{r}_m$. Thus, one  can write the total electric field of the radiation emitted from comb beam as:
\begin{eqnarray}
\textbf{E}(\textbf{r},\omega)= A(\textbf{r},\omega) \sum_{m=0}^{N_m-1}\Bigg\{\bigg(\sum_{j=1}^{N_e} \exp\Big(i\omega\dfrac{\hat{\textbf{n}}\boldsymbol{\cdot} \textbf{r}_{j,m}}{c}\Big)\bigg) \qquad \nonumber\\
 \int_{- \infty}^{+\infty} \hat{\textbf{n}} \times \Big(\hat{\textbf{n}}\times \boldsymbol{\beta_m} \Big)  \exp\bigg(i\omega \bigg(t_r -\hat{\textbf{n}}\boldsymbol{\cdot} \dfrac{\textbf{r}_c+ \boldsymbol{\Delta}\textbf{r}_m}{c}\bigg)\bigg)dt_r\Bigg\} \nonumber\\
\end{eqnarray}
As already mentioned, the separation between the micro-bunches is independent of time and therefore; the integral can be now separated from the summation:
\begin{eqnarray}
\textbf{E}(\textbf{r},\omega)= A(\textbf{r},\omega) \sum_{m=0}^{N_m-1}\Bigg\{\sum_{j=1}^{N_e} \exp\Big(i\omega \Big(\hat{\textbf{n}} \boldsymbol{\cdot} \dfrac{\textbf{r}_{j,m}+\boldsymbol{\Delta} \textbf{r}_m}{c}\Big)\Big)\Bigg\} \nonumber\\ 
\int_{- \infty}^{+\infty} \hat{\textbf{n}} \times \Big(\hat{\textbf{n}}\times \boldsymbol{\beta_m} \Big)  \exp\bigg(i\omega \bigg(t_r -\dfrac{\hat{\textbf{n}} \boldsymbol{\cdot} \textbf{r}_c}{c}\bigg)\bigg)dt_r \nonumber \\
\end{eqnarray}
 The above expression gives the general form of the electric field that can be applied to radiation emitted from a comb beam. The amplitude of the total electric field obtained from the comb electron beam can be written as:
\begin{equation}
\Big|\textbf{E}(\textbf{r},\omega)\Big|=\Big|A(\textbf{r},\omega)\Big| N_mN_e B_{avg} I(t)
\end{equation}
where
\begin{eqnarray}
B_{avg}\big(\omega,\dfrac{\boldsymbol{\Delta}\textbf{r}_m}{c}\big)=\dfrac{\bigg| \displaystyle \sum_{m=0}^{N_m-1}\sum_{j=1}^{N_e} \exp\Big(\dfrac{i\omega\hat{\textbf{n}} \boldsymbol{\cdot} \left(\textbf{r}_{j,m}+\boldsymbol{\Delta} \textbf{r}_m\right)}{c}\Big)\bigg|}{N_m N_e}\nonumber\\
\end{eqnarray}

is called the average bunching factor of the comb beam. Thus, in the case of the comb beam, we can define bunching factor in two ways i.e. the individual bunching factor of a micro-bunch given by $B(\omega)$ and the average bunching factor of the comb beam distribution given by $B_{avg}$. 
The above definition of the bunching factor leads to following two remarks:\\
1. If the bunch structure of all the micro-bunches is exactly the same and the individual bunching factor has the maximum value i.e. 1, then the average bunching factor can be written as
\begin{eqnarray} \label{BF_Comb}
B_{avg}\big(\omega,\dfrac{\boldsymbol{\Delta}\textbf{r}_m}{c}\big) =\dfrac{\Bigg| \displaystyle \sum_{m=0}^{N_m-1}\exp(i\omega\hat{\textbf{n}}\boldsymbol{\cdot}\boldsymbol{\Delta}\textbf{r}_m/c)\Bigg|}{N_m} \nonumber\\ =\Bigg|\sum_{m=0}^{N_m-1} \exp⁡\Big(i\omega\dfrac{ \hat{\textbf{n}}\boldsymbol{\cdot}\boldsymbol{\Delta}\hat{\textbf{r}}_m }{\omega_{c}}\Big)\Bigg| \Big/N_m \nonumber \\
\end{eqnarray}
\begin{figure*}
	\includegraphics[width=17.5cm,keepaspectratio]{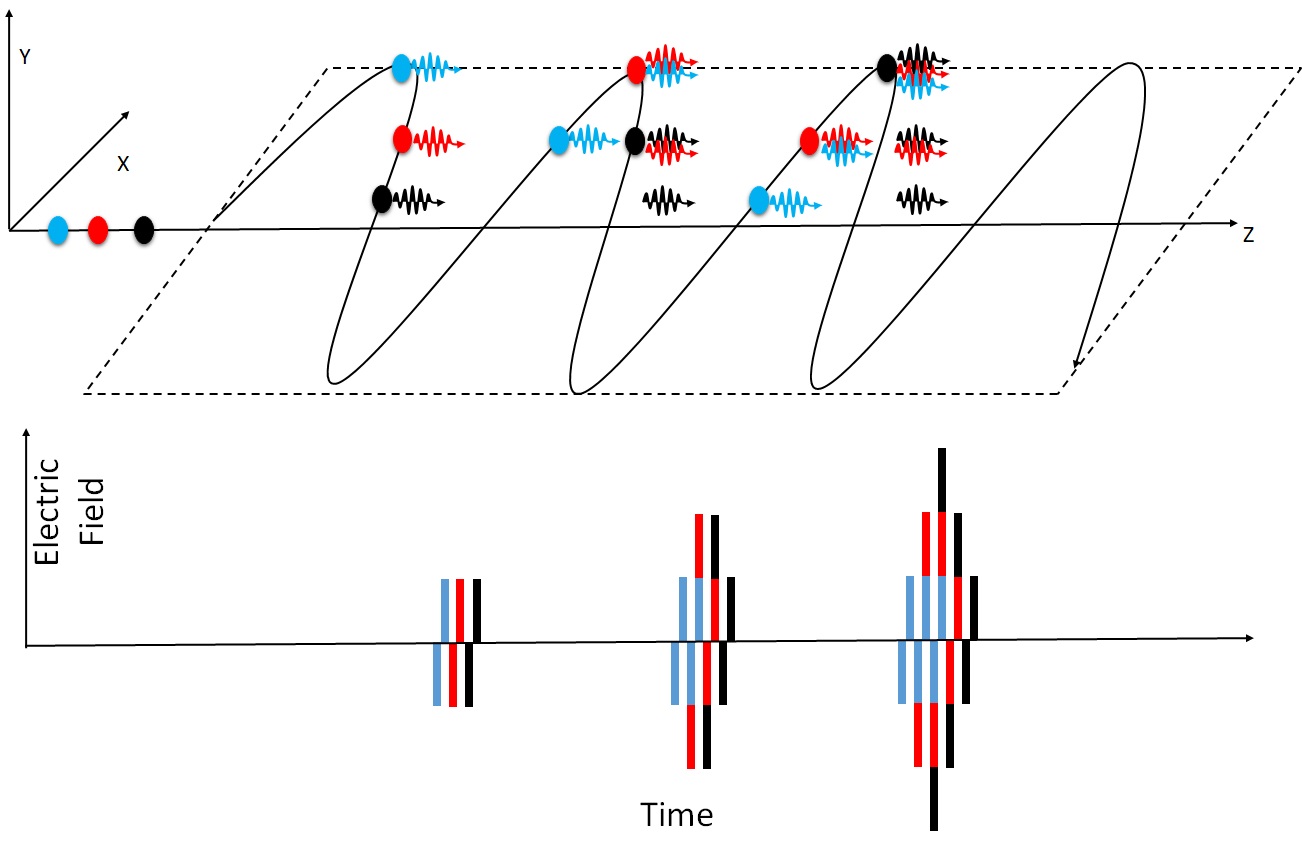}
	\captionof{figure}{Schematic of constructive interference of the radiation wave packets emitted from different micro-bunches of the comb beam as they travel inside the undulator (Top) and the increase in Electric Field of the radiation over different undulator periods (Bottom).}{\label{fig:dls_schematic} }
\end{figure*}
where $\omega_{c}=c/\Delta r_m $ is the frequency corresponding to the temporal separation between the micro-bunches of the comb beam. Eq. \ref{BF_Comb} shows that the average bunching factor will depend on the relative separation between any two micro-bunches. If the micro-bunches are distributed in a way that $\Delta r_m$ points towards the direction of the observer (located along the axis of the undulator), then the bunching factor will be maximum only if the separation between the micro-bunches is exactly equal to one radiation wavelength i.e. $\Delta r_m=m\lambda$. If this condition is satisfied, then the electric field of the radiation mode $\omega$ will be proportional to $N_m N_e$  and the radiation intensity emitted into the radiation mode will scale as $(N_m N_e )^2$. An example of such a distribution would be the electron beam distribution at the saturation of the Self-Amplified Spontaneous Emission (SASE) FEL in which the injected electron beam interacts continuously with the radiation field and gets micro-bunched (if injected at the correct phase of the ponderomotive potential) such that the individual micro-bunches are super-radiant and are separated by one radiation wavelength \cite{Tremaine2002,Huang2007}. \\
2. If $\omega_{c}$  is exactly equal to the fundamental frequency of the undulator $\omega$, then the average bunching factor will depend on the structure of the individual micro-bunches. The micro-bunches having lower individual bunching factor will emit incoherent or partially coherent radiation; while the micro-bunches that have higher individual bunching factor will emit coherent radiation. \\
3. An important comparison can be drawn from the definition of the bunching factor of a single bunch $B_{single}$ and that of a train of micro-bunches ($B_{comb}$); calculated for a particular frequency $\omega$. The bunching factor of the single bunch will always be greater than or equal to $B_{single}$ for frequencies lower than $\omega$; however; this is not true for the comb beam. The average bunching factor of the comb beam will peak at  $\omega_c$ and its harmonics; while it will quickly drop to values $\sim$0 for all other frequencies.\\
In Section 3, the individual and average bunching factors have been computed for different cases of comb beam and their effect on the radiation output is discussed in Section 4.\par
In our project, Delhi Light Source, the efforts will be made to produce comb electron beam in which individual micro-bunches are super-radiant and the separation between the micro-bunches of the comb beam will be tuned to the resonant frequency of the undulator so that the individual and the average bunching factor is as high as possible. It is then expected that the radiation wave packet emitted from the trailing micro-bunches will slip ahead of their respective sources (by one radiation wavelength every undulator period) and constructively interfere with the radiation wave packets emitted from the leading micro-bunches (as shown in Fig.\ref{fig:dls_schematic}). This will result in a linear growth in the electric field amplitude; which will get saturated after $N_m$ number of undulator periods.\par

\section{\label{BeamOptics}Beam Optics Simulations}
\begin{figure*}
\includegraphics[height=5.7cm,keepaspectratio]{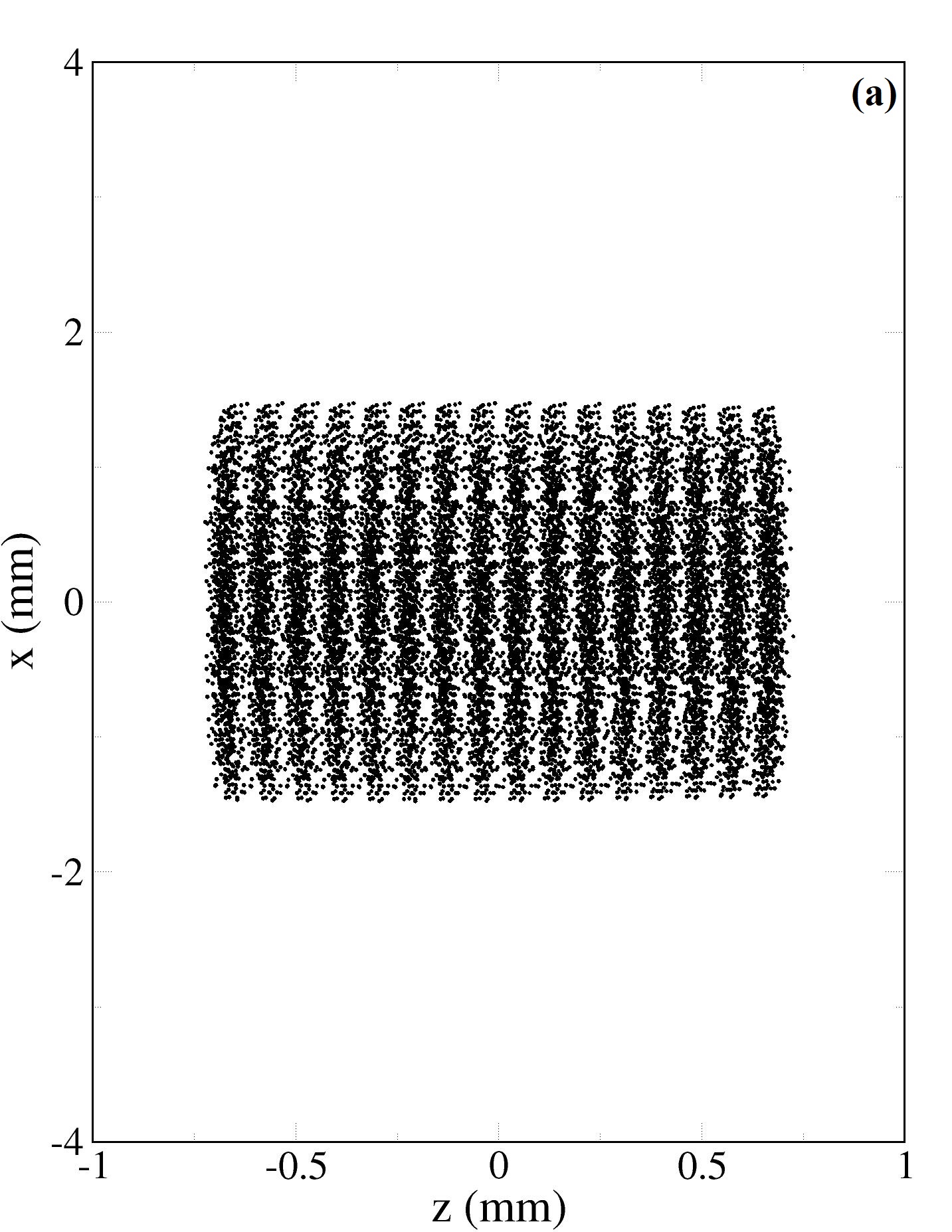}\includegraphics[height=5.7cm,keepaspectratio]{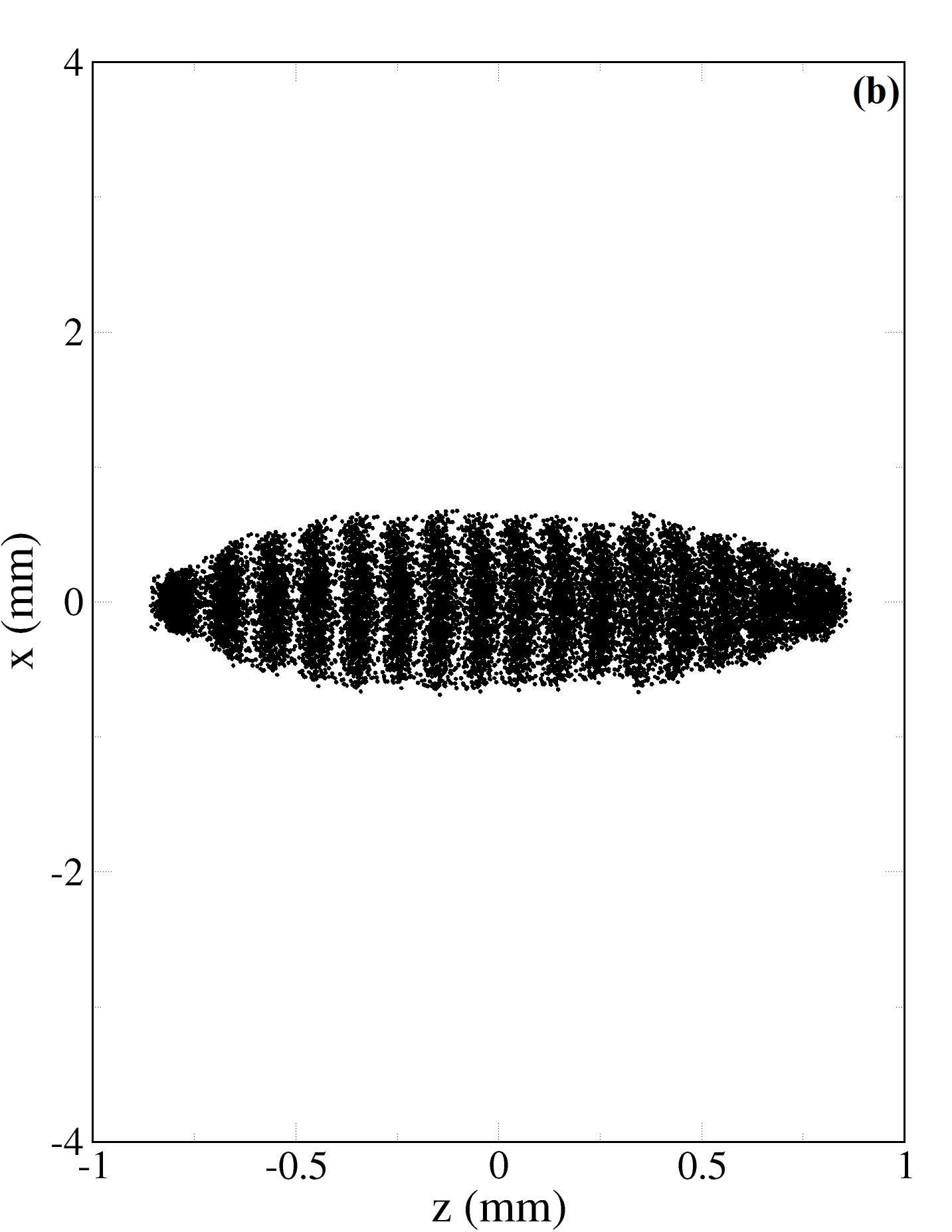}\includegraphics[height=5.7cm,keepaspectratio]{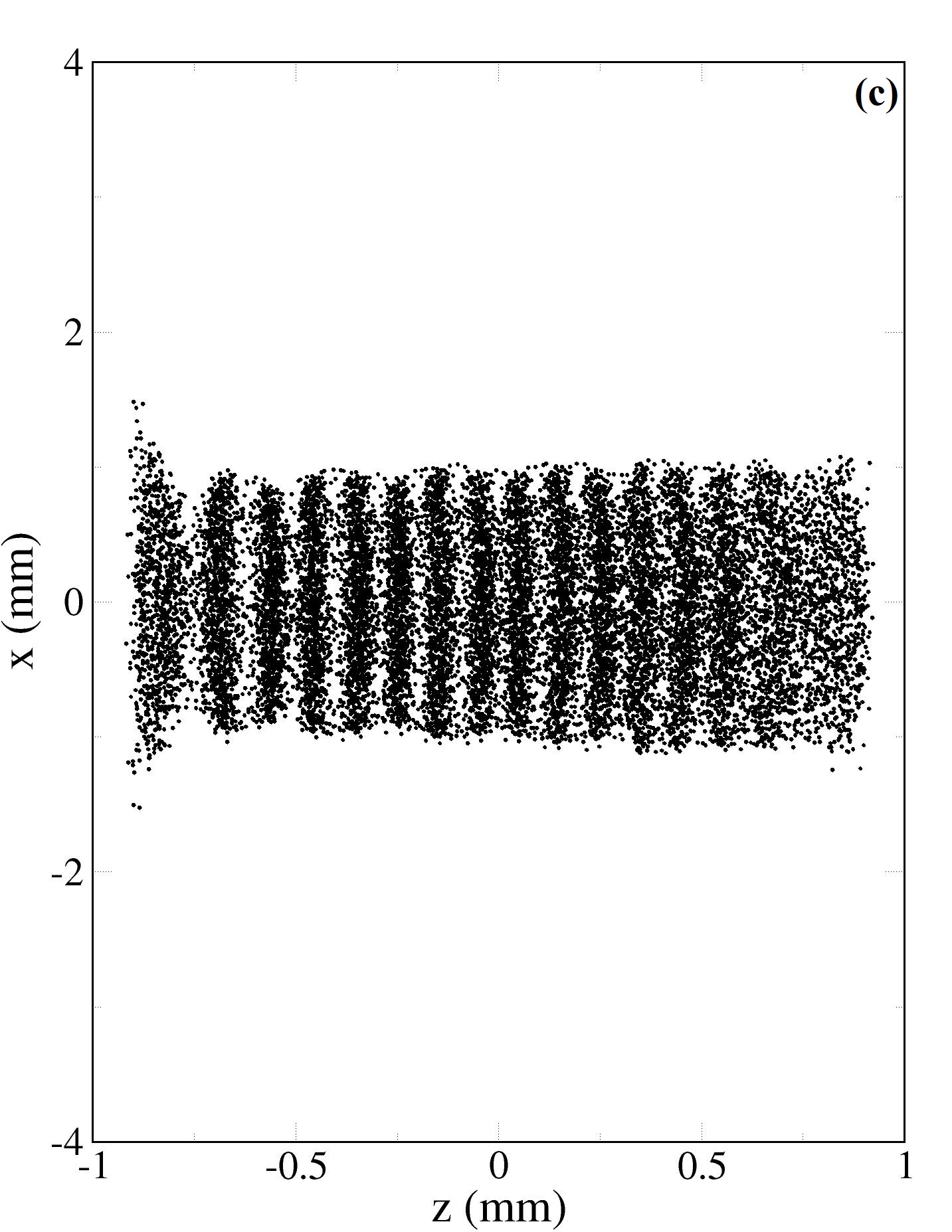}\includegraphics[height=5.7cm,keepaspectratio]{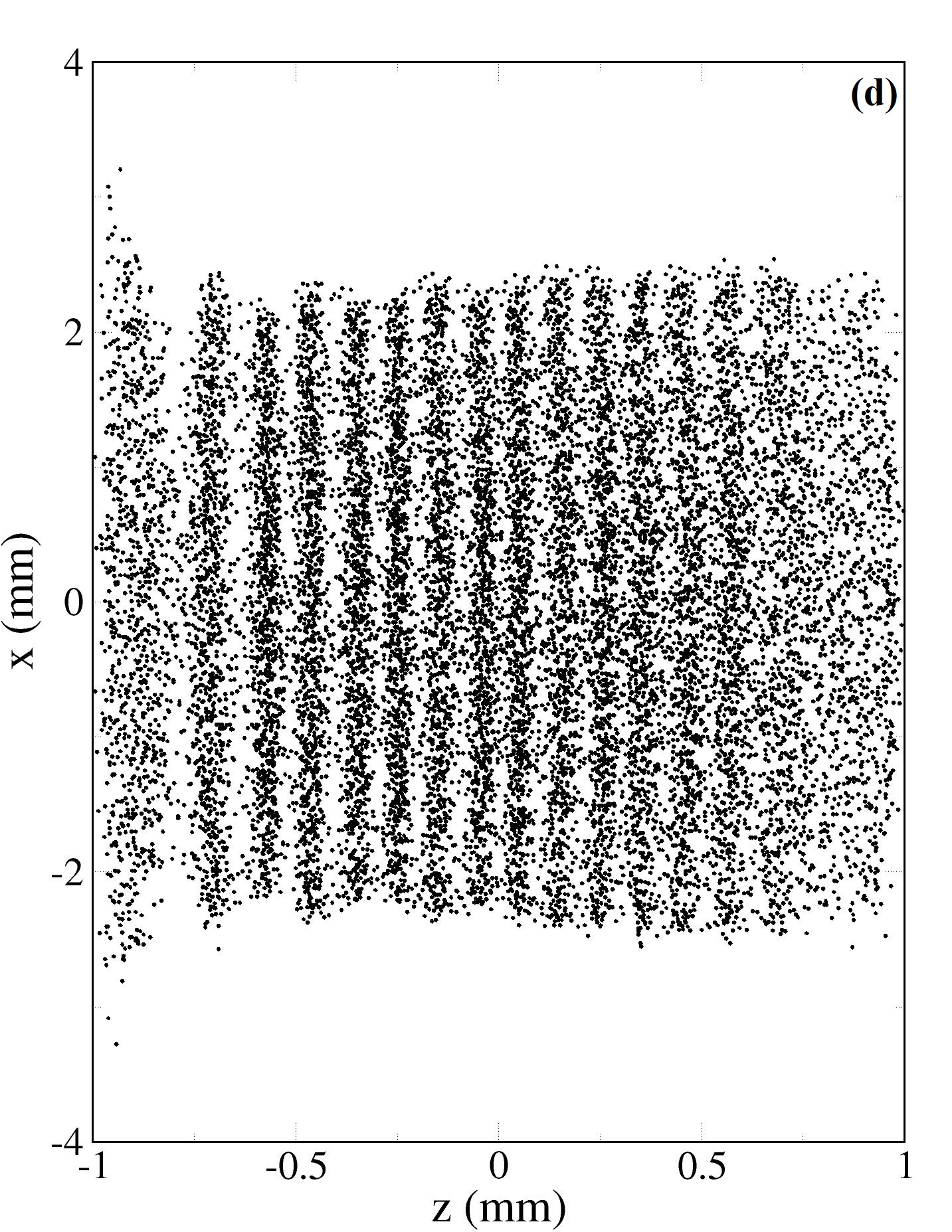}\\
\includegraphics[height=5.72cm,keepaspectratio]{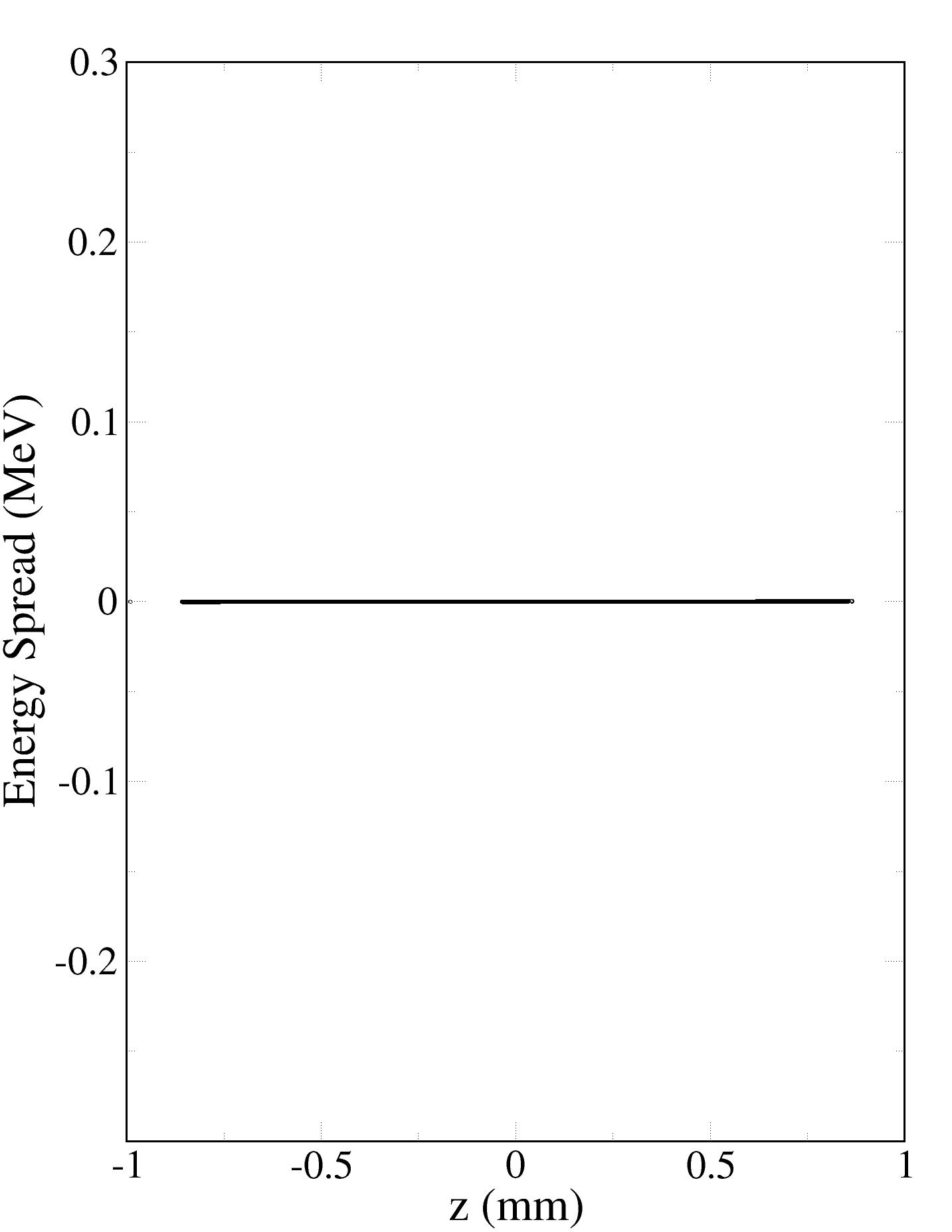}\includegraphics[height=5.72cm,keepaspectratio]{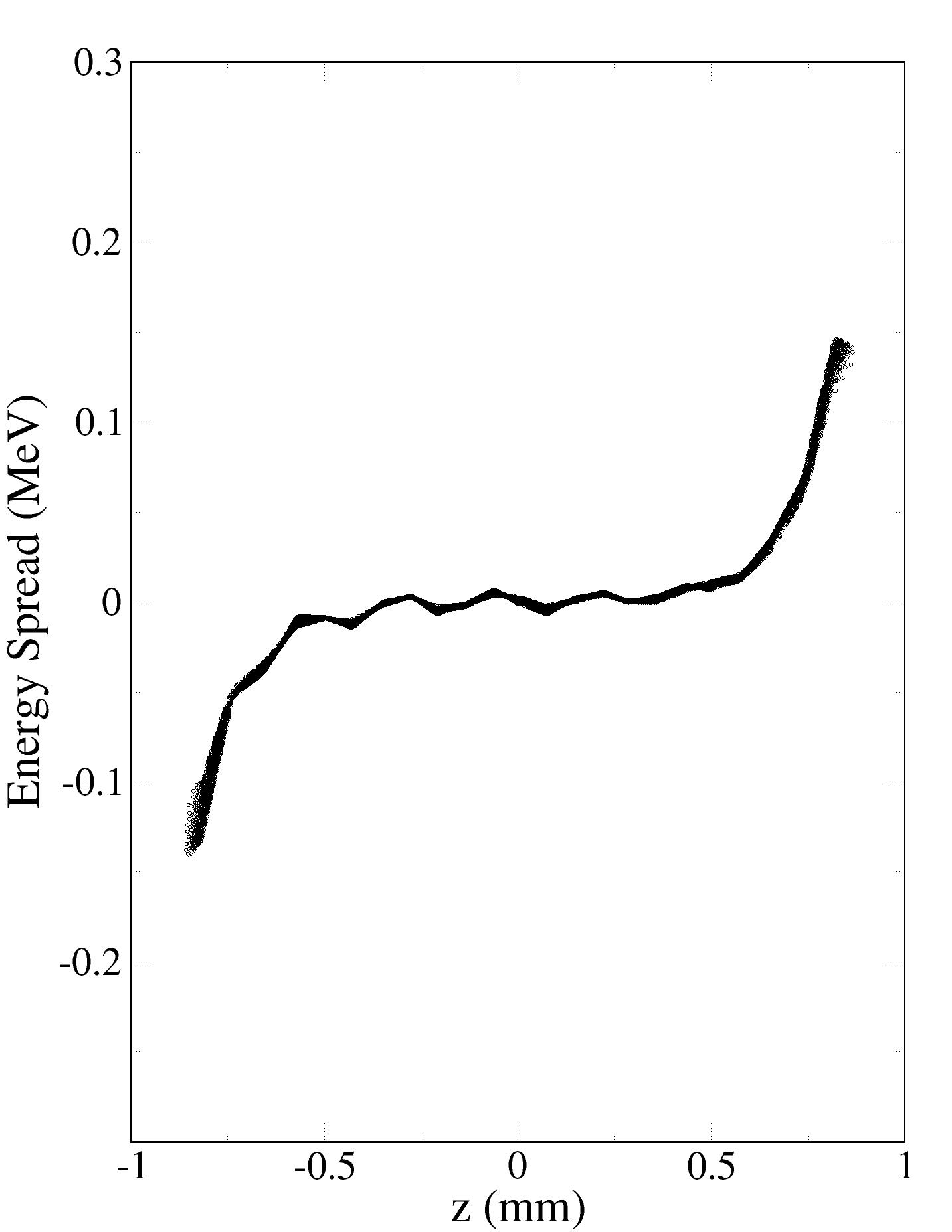}\includegraphics[height=5.72cm,keepaspectratio]{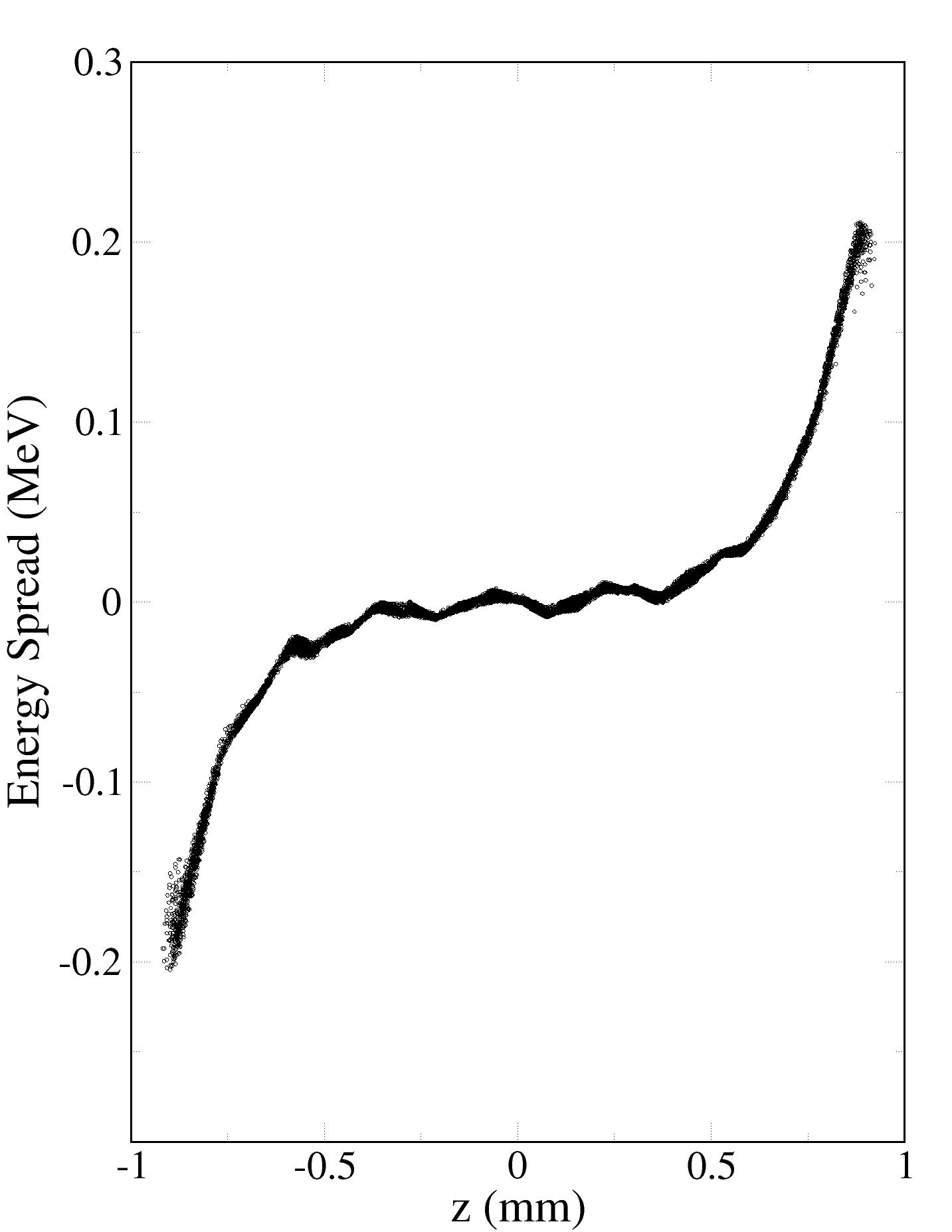}\includegraphics[height=5.72cm,keepaspectratio]{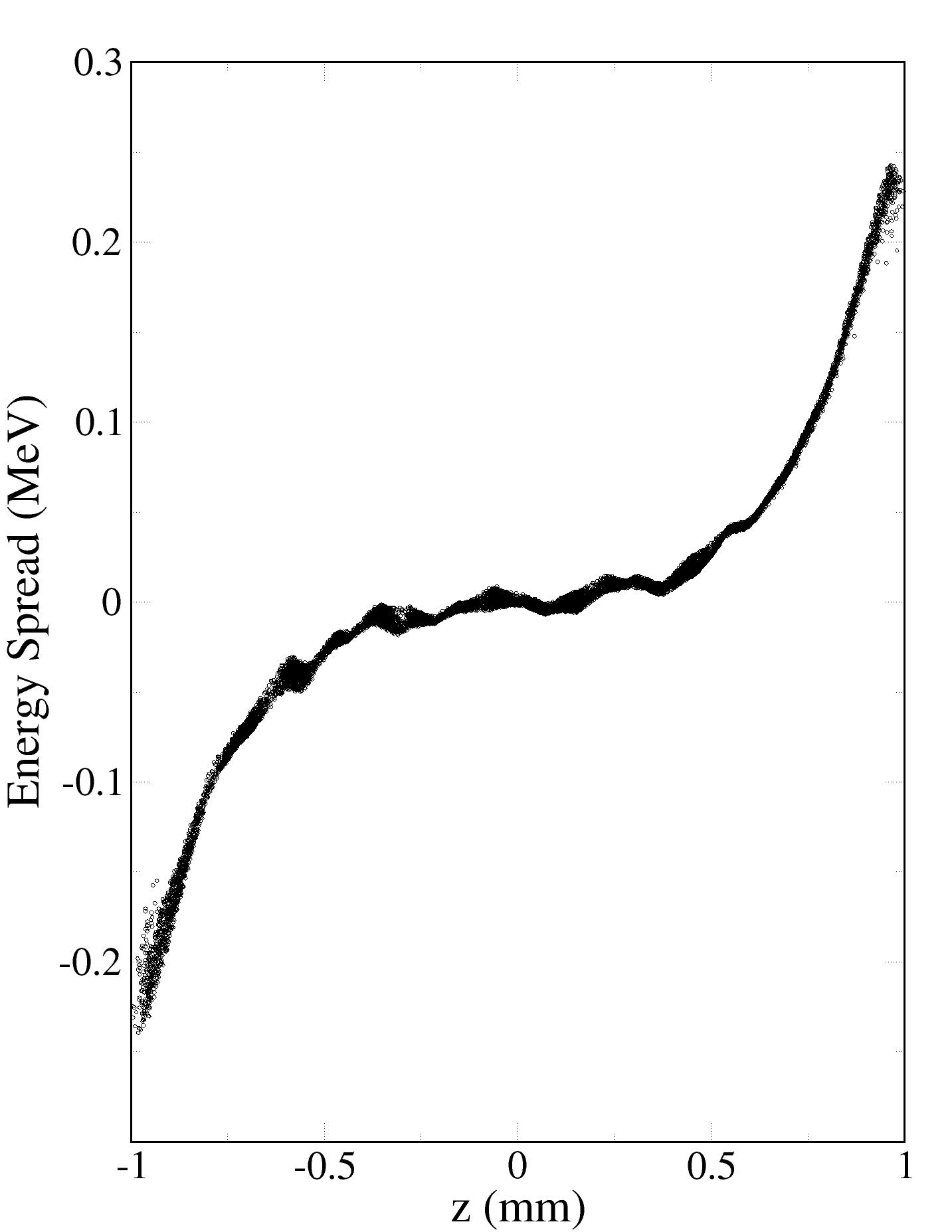}\\
\includegraphics[height=5.7cm,keepaspectratio]{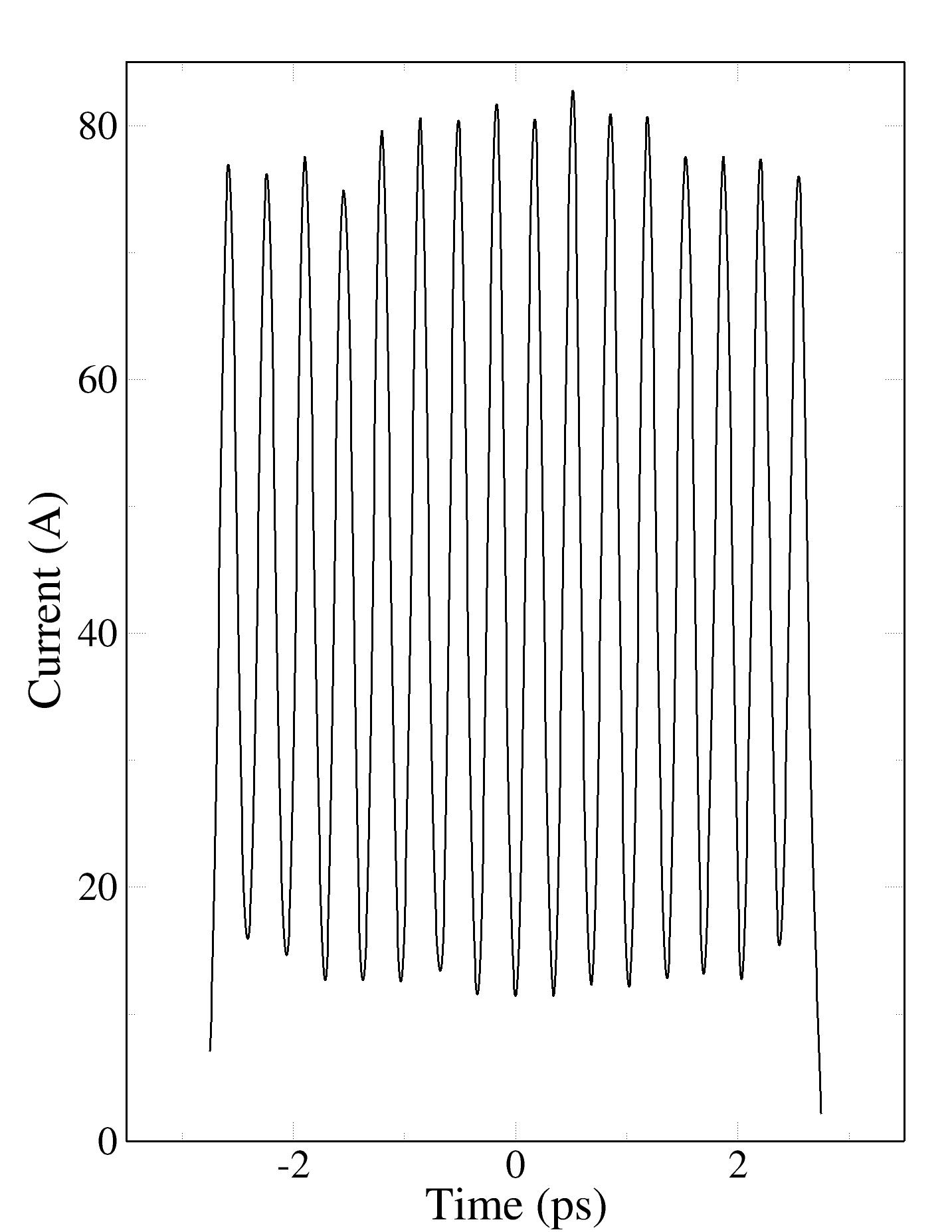}\includegraphics[height=5.7cm,keepaspectratio]{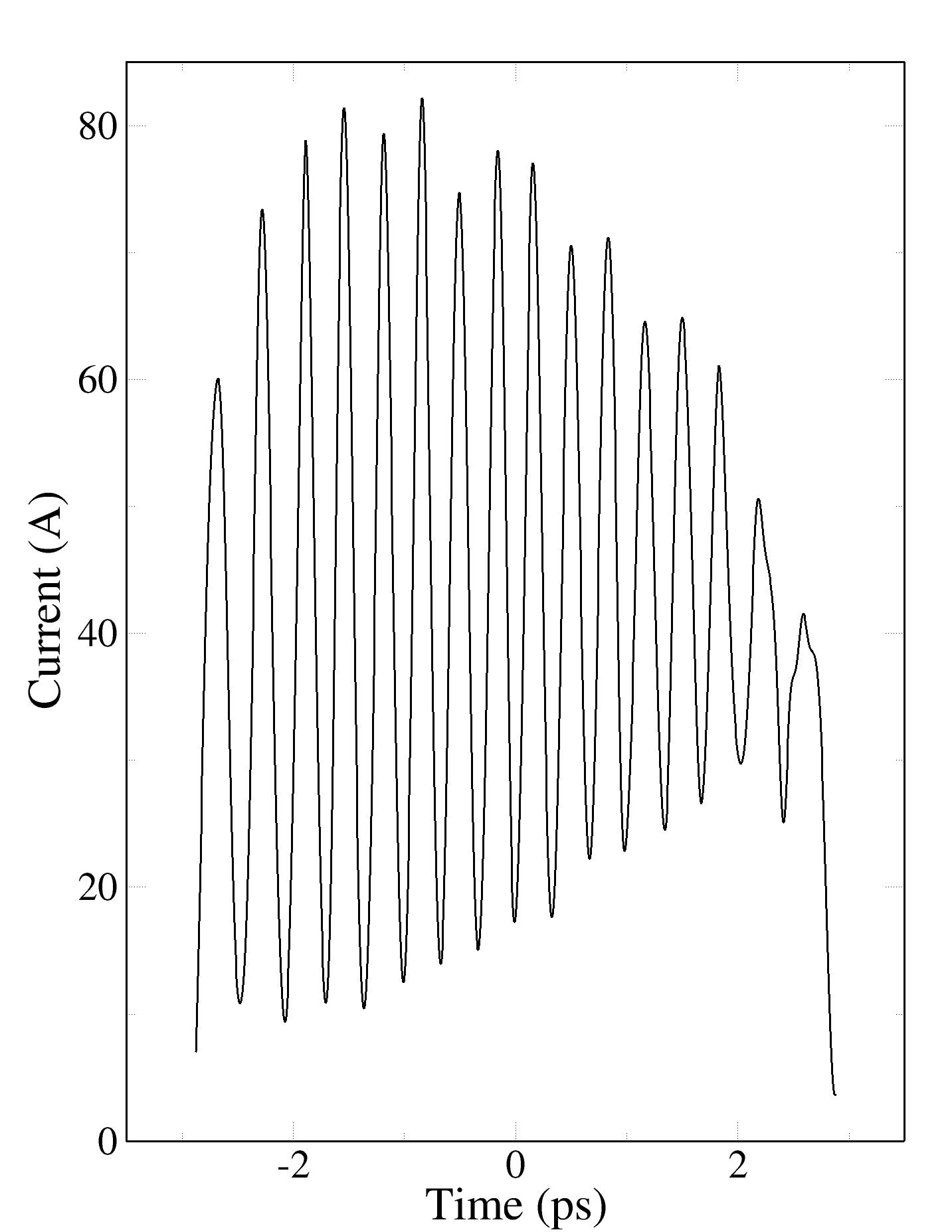}\includegraphics[height=5.7cm,keepaspectratio]{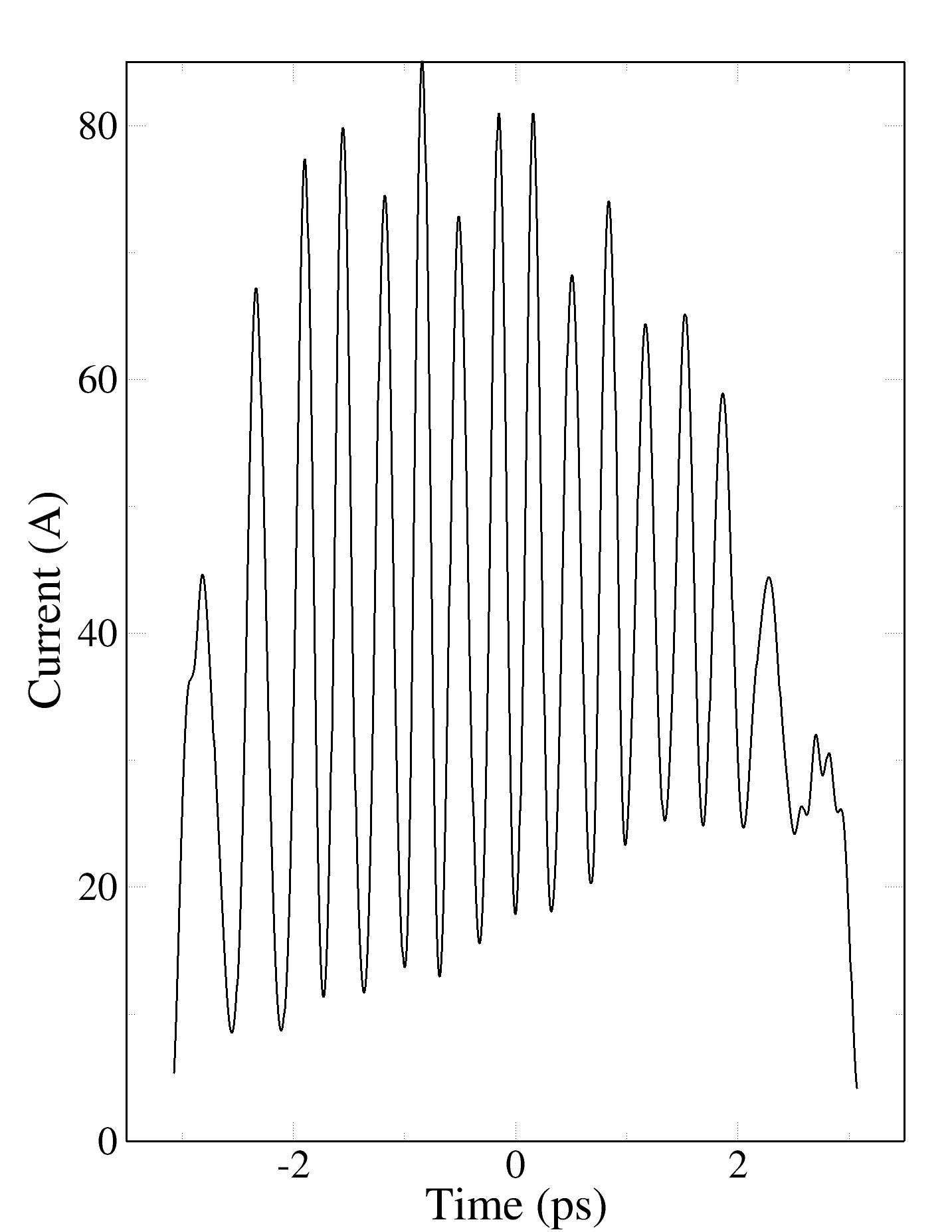}\includegraphics[height=5.7cm,keepaspectratio]{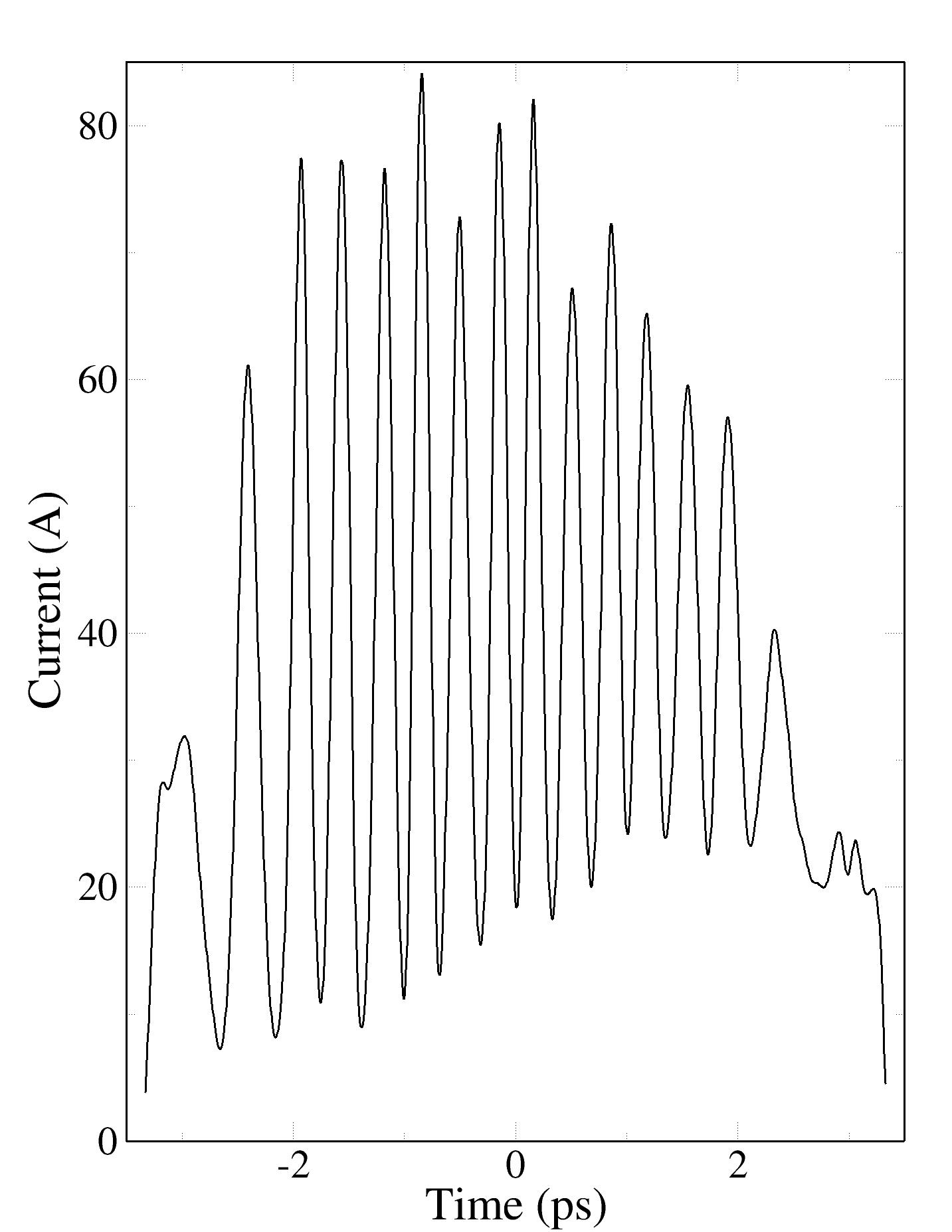}
\caption{\label{fig:3THz_beamoptics}Snapshot of electron beam profile (Top), Energy Spread (Middle) and Peak Current Distribution (Bottom) for 3 THz at (a) Photocathode (b) Undulator's entrance (c) Undulator's centre and (d) Undulator's exit.}
\includegraphics[width=5.43cm,keepaspectratio]{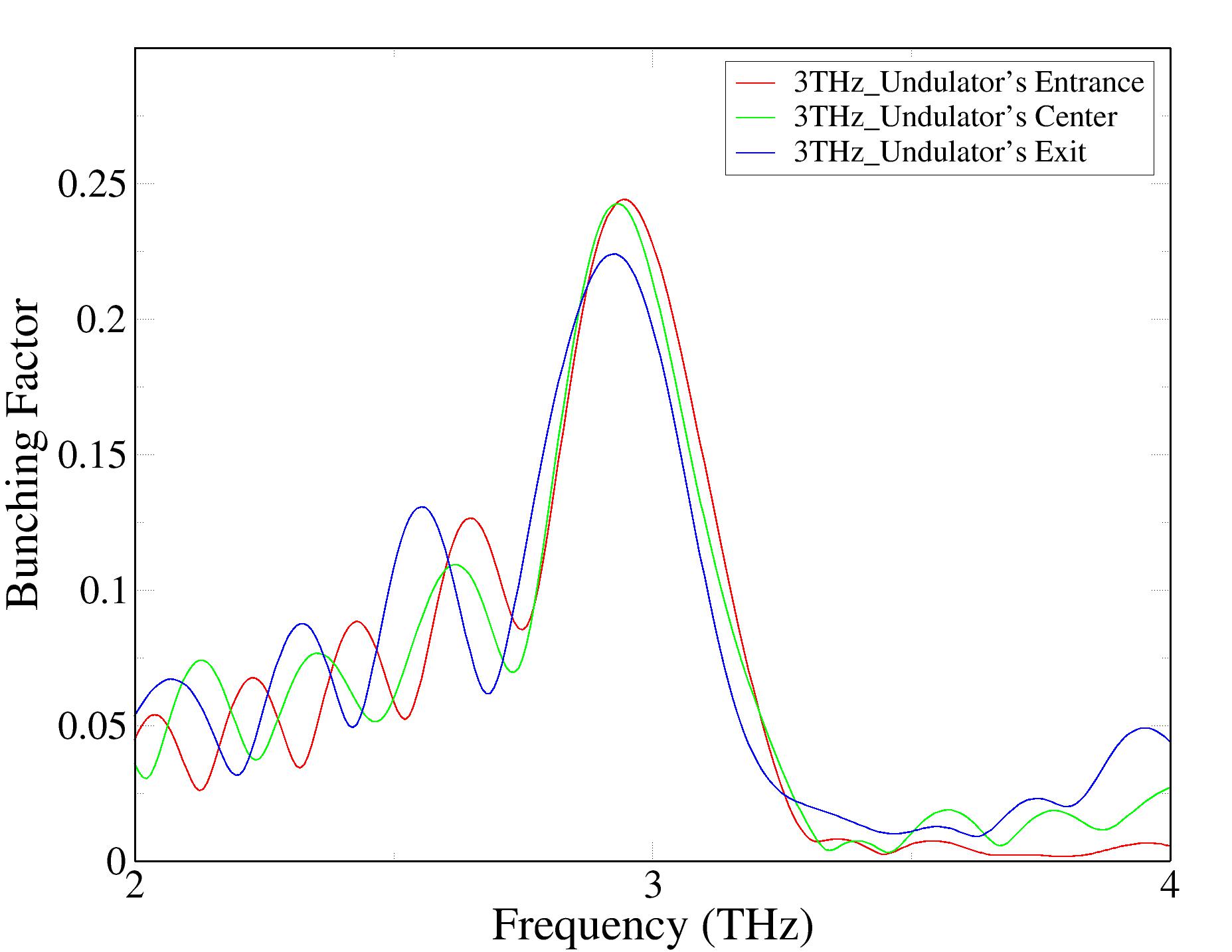}\includegraphics[width=5.43cm,keepaspectratio]{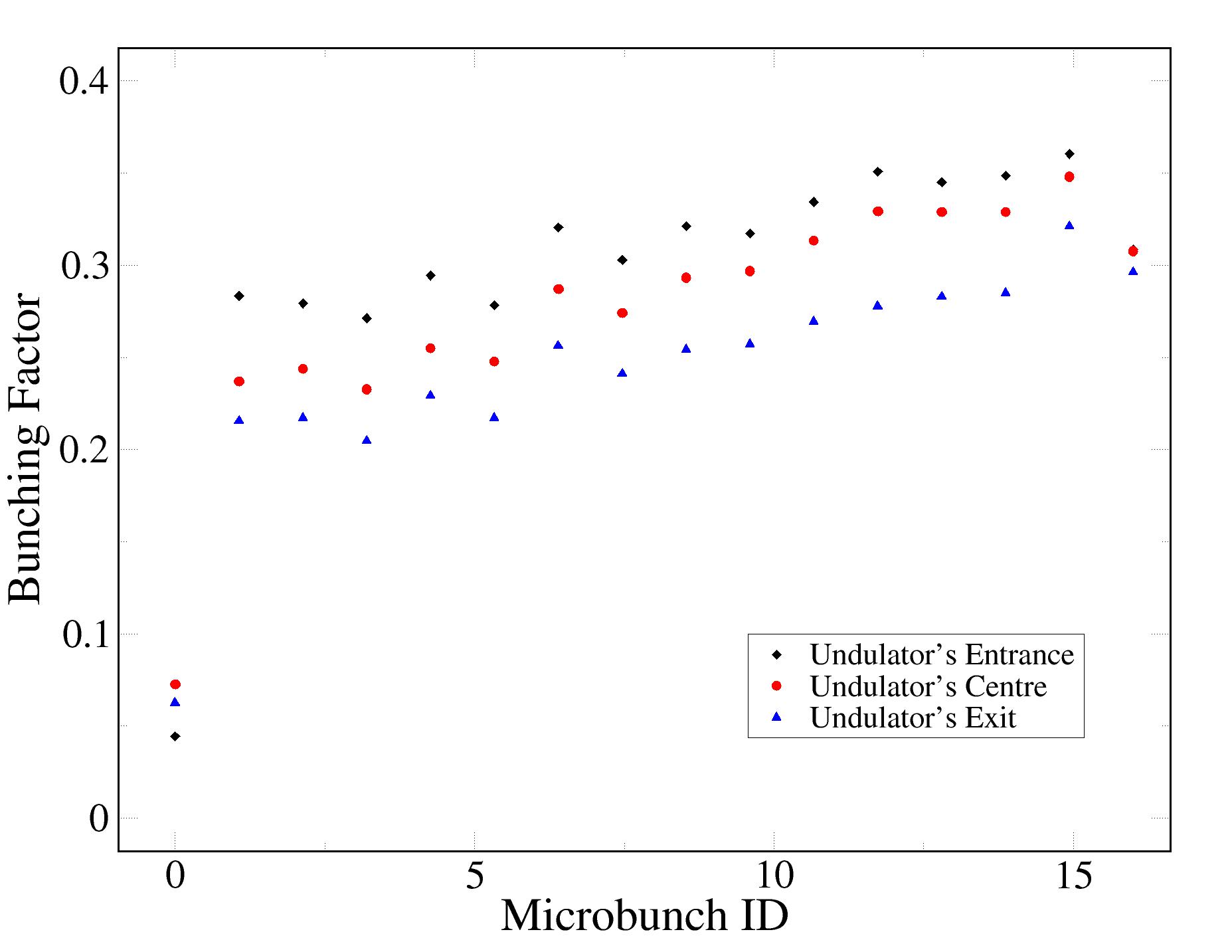}\includegraphics[width=5.43cm,keepaspectratio]{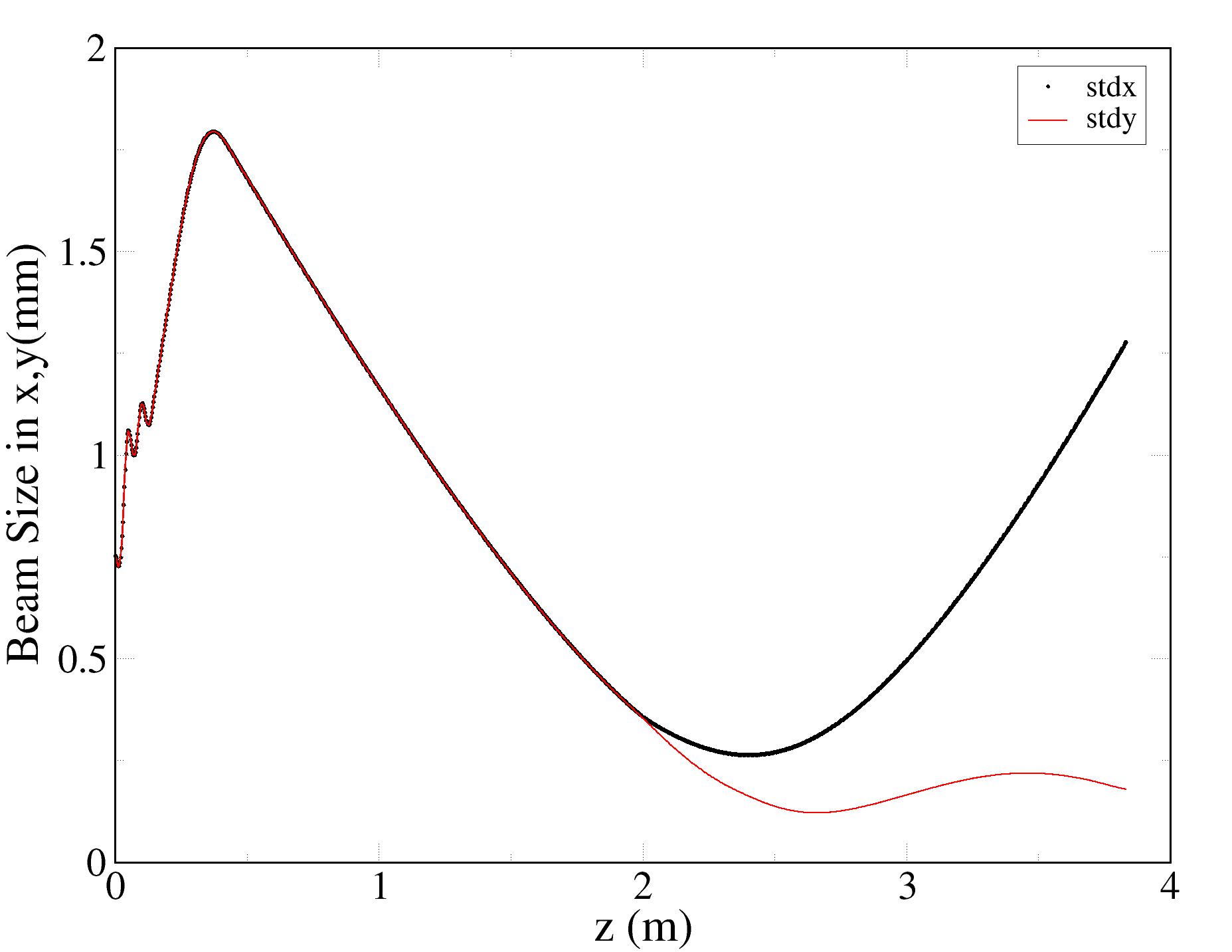}
\captionof{figure}{\label{fig:dls_3THZ_bf}Variation of the comb beam average bunching factor (left), individual bunching factor (middle) and envelope through the beamline for 3 THz (right).}
\end{figure*}

The beam optics calculations were performed using the GPT code to check the feasibility of generating the comb electron beam by laser micro-pulses incident on the photocathode \cite{Ghosh2017,Aryshev2017}. The electron beam generated at the photocathode was accelerated by the 2.6 cell RF cavity and subsequently focused by the solenoid and the quadrupole magnet to produce a very tight beam size in y-direction (a few hundreds of micron) and somewhat larger size in x-direction (a few m.m.) (Fig. \ref{fig:3THz_beamoptics}(right)). The main objective of the beam optics simulation is to transport the beam from the photocathode to the exit of the undulator magnet (see Fig. \ref{fig:dls_beamline}) by ensuring (a) minimum size along y-axis (b) minimum energy spread $\Delta$E (c) minimum emittance (d) maximum average bunching factor of the electron beam and (e) maximum individual bunching factor of the micro-bunches in the comb beam bunches throughout the length of the undulator. Other important point to consider is the dispersion of the electron beam inside the undulator due to energy chirp in the multi micro-bunch structure\cite{Campbell2015,Prat2010}. The energy chirp in the pre-bunched electron beam will result in different transverse trajectories of the particles inside the undulator causing de-bunching of the micro-bunched beam. It is expected that different micro-bunches in electron beam will have different energies as well as different energy spread as they are produced at different RF phases of the electron gun \cite{Boscolo2007}.  For the increased energy spread of the micro-bunches, longitudinal de-bunching due to dispersion will be larger resulting in a drop in bunching factor.\par The beam optics calculation has been done for 3 THz and 0.6 THz with radial laser beam and major simulation parameters are given in Table \ref{tab:table1}.
For the case of 3 THz, Fig. \ref{fig:3THz_beamoptics} shows the longitudinal profile, energy spread and the current of the comb beam at the photocathode and at different positions inside the undulator. It is evident from Fig. \ref{fig:3THz_beamoptics} that while the beam traverses through the beamline to reach the undulator's entrance, the last and the first few micro-bunches (out of the 16 micro-bunches) develop significant energy spread ($\Delta E$ $\sim$100 keV) and have a tendency to merge with each other. Due to large longitudinal space charge forces acting from opposite directions, the micro-bunches located well within the bunch are able to maintain their longitudinal bunching and only a small change in their energy spread and peak current (80 Amperes) are observed. It is also observed that except for the first micro-bunch; all the other micro-bunches have individual bunching factor greater than 0.2 and the average bunching factor is maintained well above 0.2 throughout the beamline (Fig. \ref{fig:dls_3THZ_bf}). Fig. \ref{fig:dls_3THZ_bf} (right) shows the evolution of the electron beam envelope ($\sigma_x$ \& $\sigma_y$)  from the photocathode to the exit of the undulator. Inside the undulator, the beam envelope along y axis is almost constant (\textless 0.25 mm) while it is allowed to expand along the x axis. The maximum possible size of the beam along x axis is limited by the drop in the magnetic field of the undulator due to the ‘magnetic roll-off’ effect which becomes significant beyond x= $\pm$ 10 mm \cite{Tripathi2017}.

\begin{table}[h]
	\caption{\label{tab:table1}%
		Comb electron beam parameters for DLS
	}
	\begin{ruledtabular}
		\begin{tabular}{llcc}
			\textrm{Frequency}&
			\textrm{THz}&
			\textrm{0.6}&
			\textrm{3}\\
			\colrule
			No. of micro-bunches&&4&	16\\
			Charge/micro-bunch &pC&	15&	15\\
			Accelerating Field &MV/m&	91&	110\\
			RF Launching Phase &degrees&	30&	30\\
			Energy &MeV&6.65&8.2\\
			Energy Spread&\%&0.31&0.6\\
			FWHM @ Photocathode &fs&200&200\\
			FWHM @ Undulator$'$s Entrance &fs&250& 200\\
			Avg. Sep. @ Undulator$'$s Entrance&fs &1630 & 335\\
			$\sigma_{x}$ &$\mu$m&400&275\\
			$\sigma_{y}$ &$\mu$m&250&175\\			
			Normalised $\epsilon_{x}$ &mm-mrad&0.80&0.65\\
			Normalised $\epsilon_{y}$& mm-mrad&0.65& 0.7
		\end{tabular}
	\end{ruledtabular}
\end{table}

\begin{figure*}
\includegraphics[height=5.7cm,keepaspectratio]{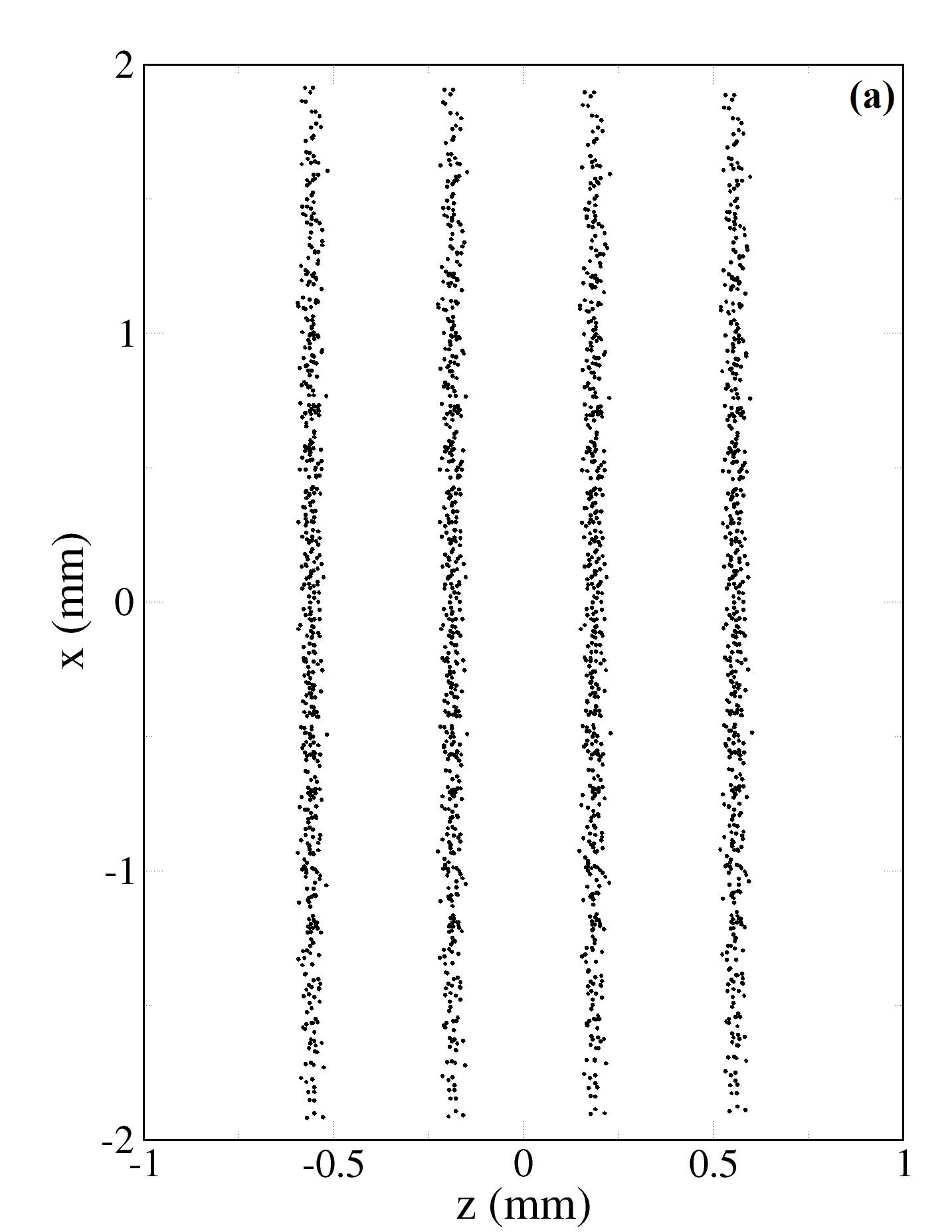}\includegraphics[height=5.7cm,keepaspectratio]{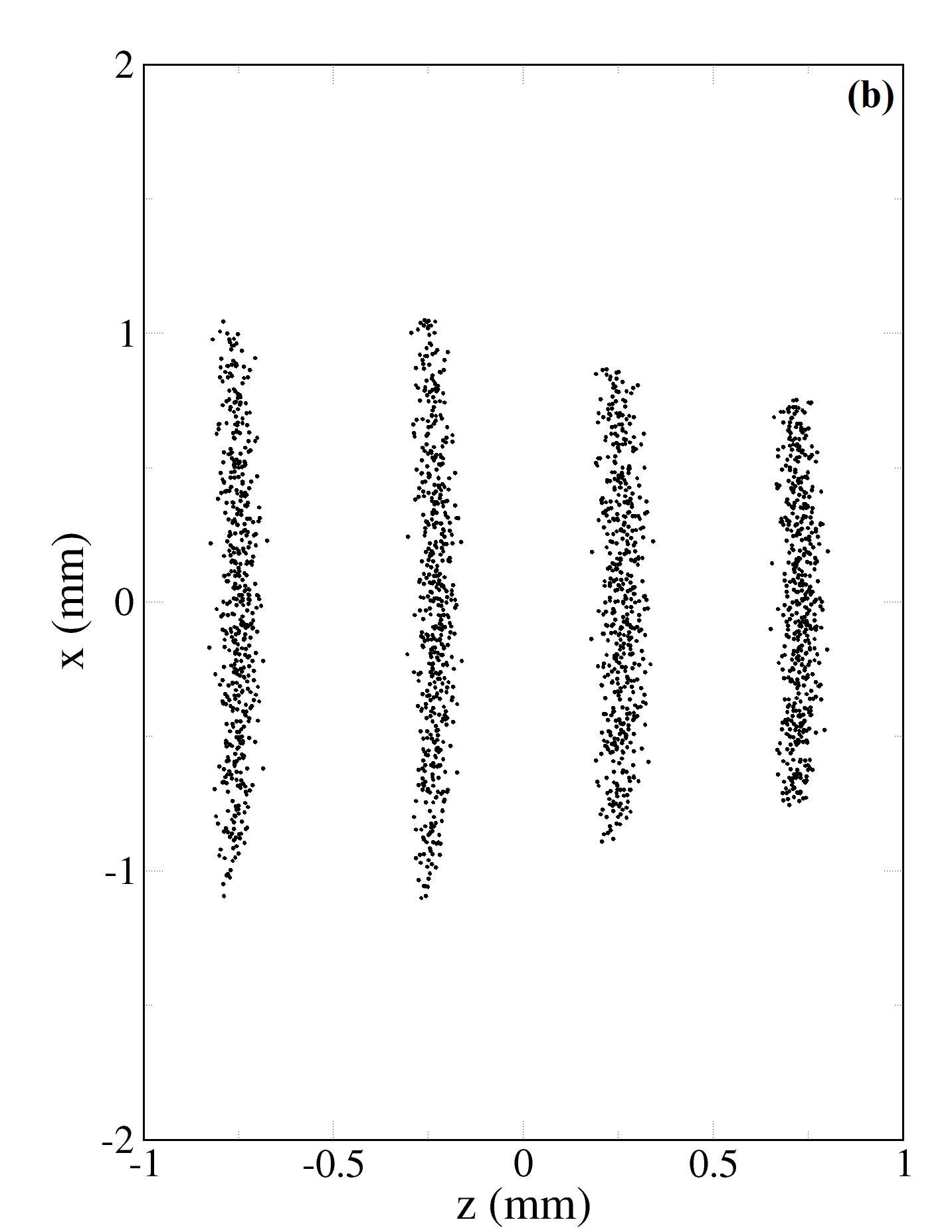}\includegraphics[height=5.7cm,keepaspectratio]{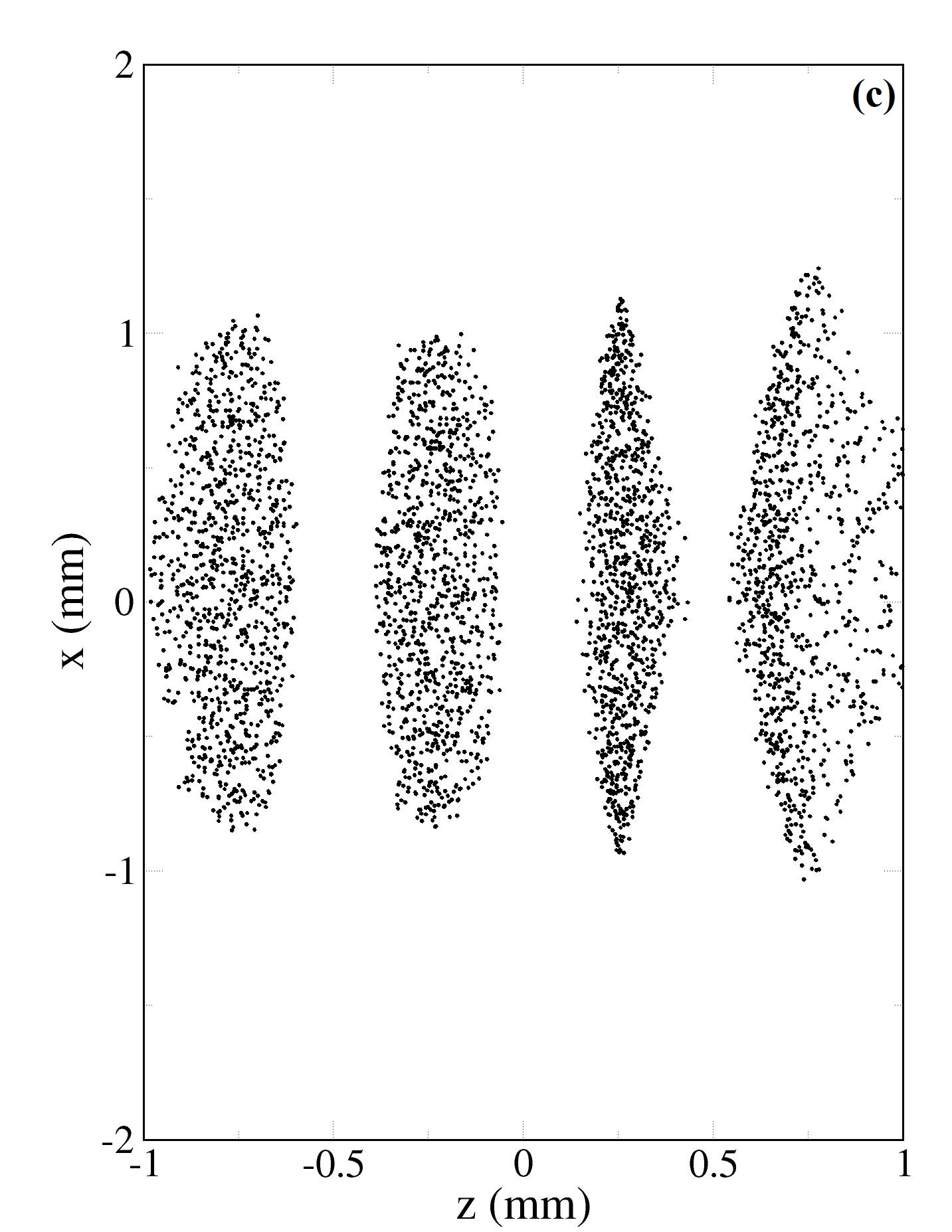}\includegraphics[height=5.7cm,keepaspectratio]{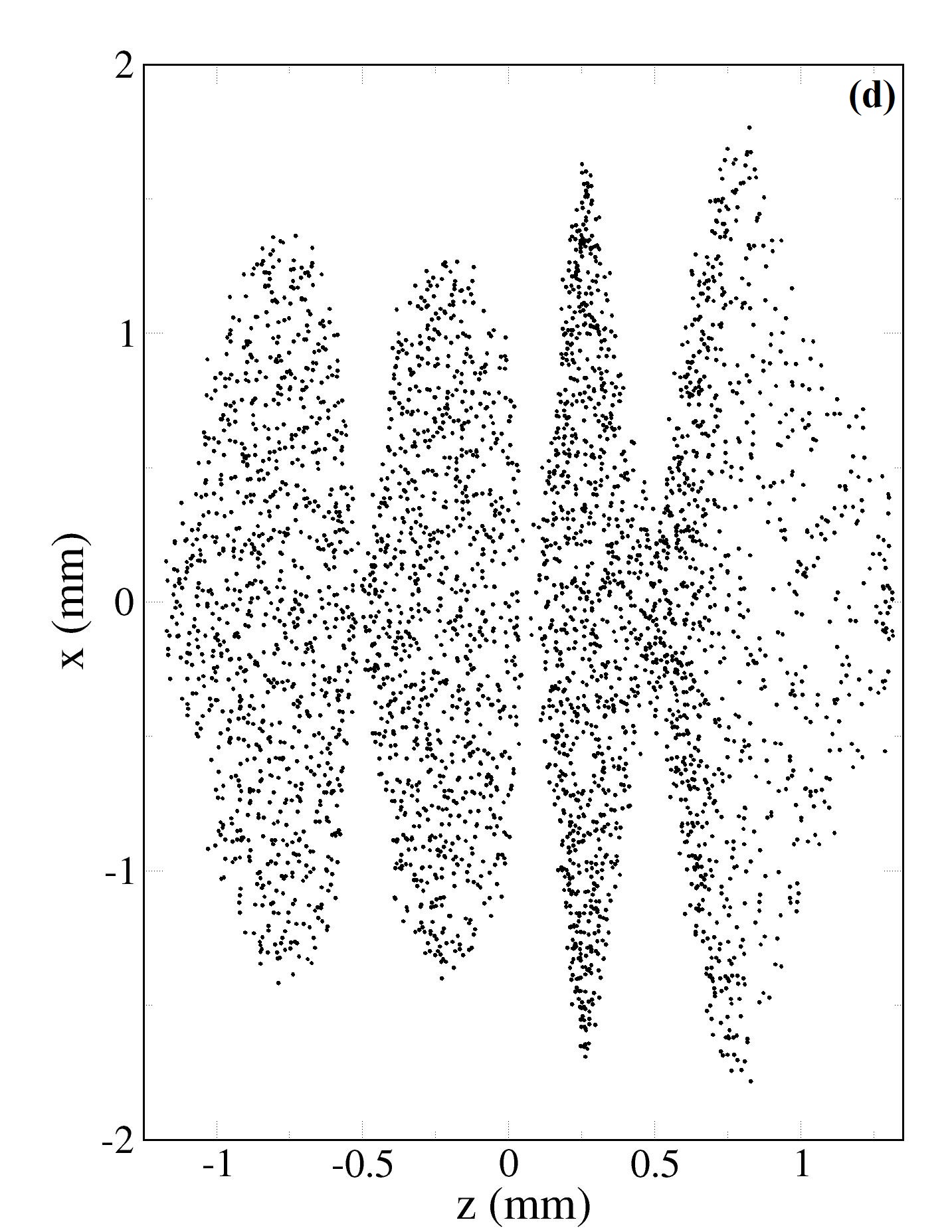}\\
\includegraphics[height=5.7cm,keepaspectratio]{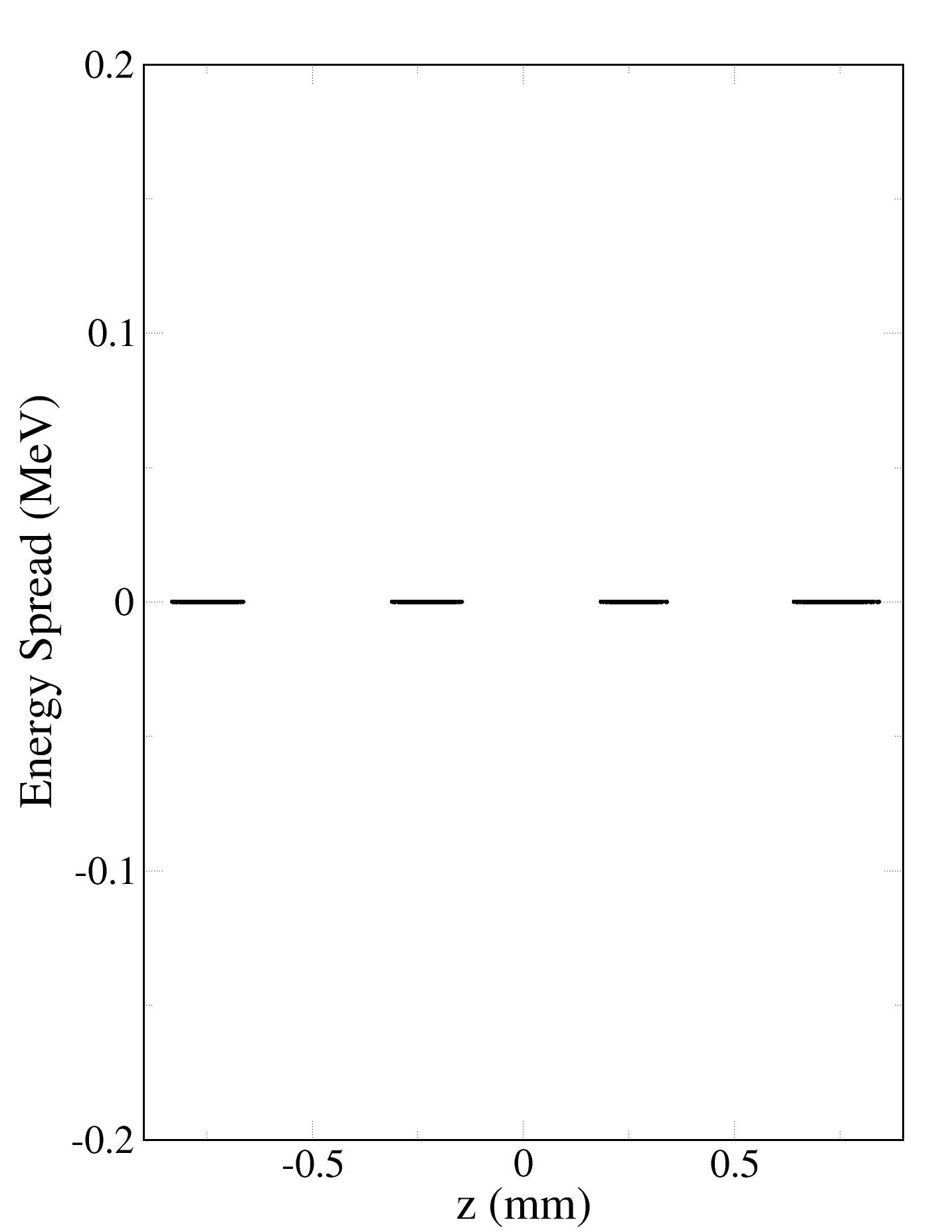}\includegraphics[height=5.7cm,keepaspectratio]{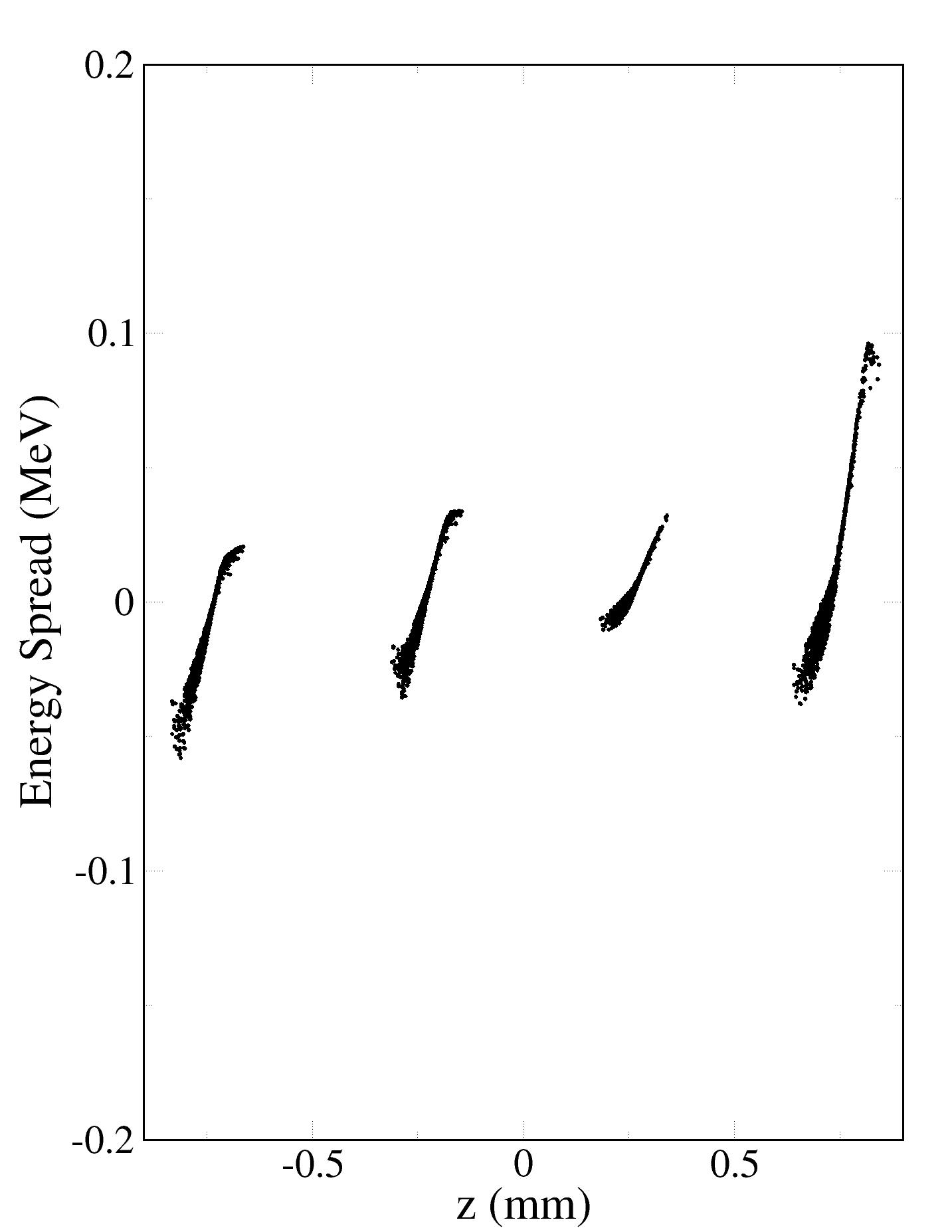}\includegraphics[height=5.7cm,keepaspectratio]{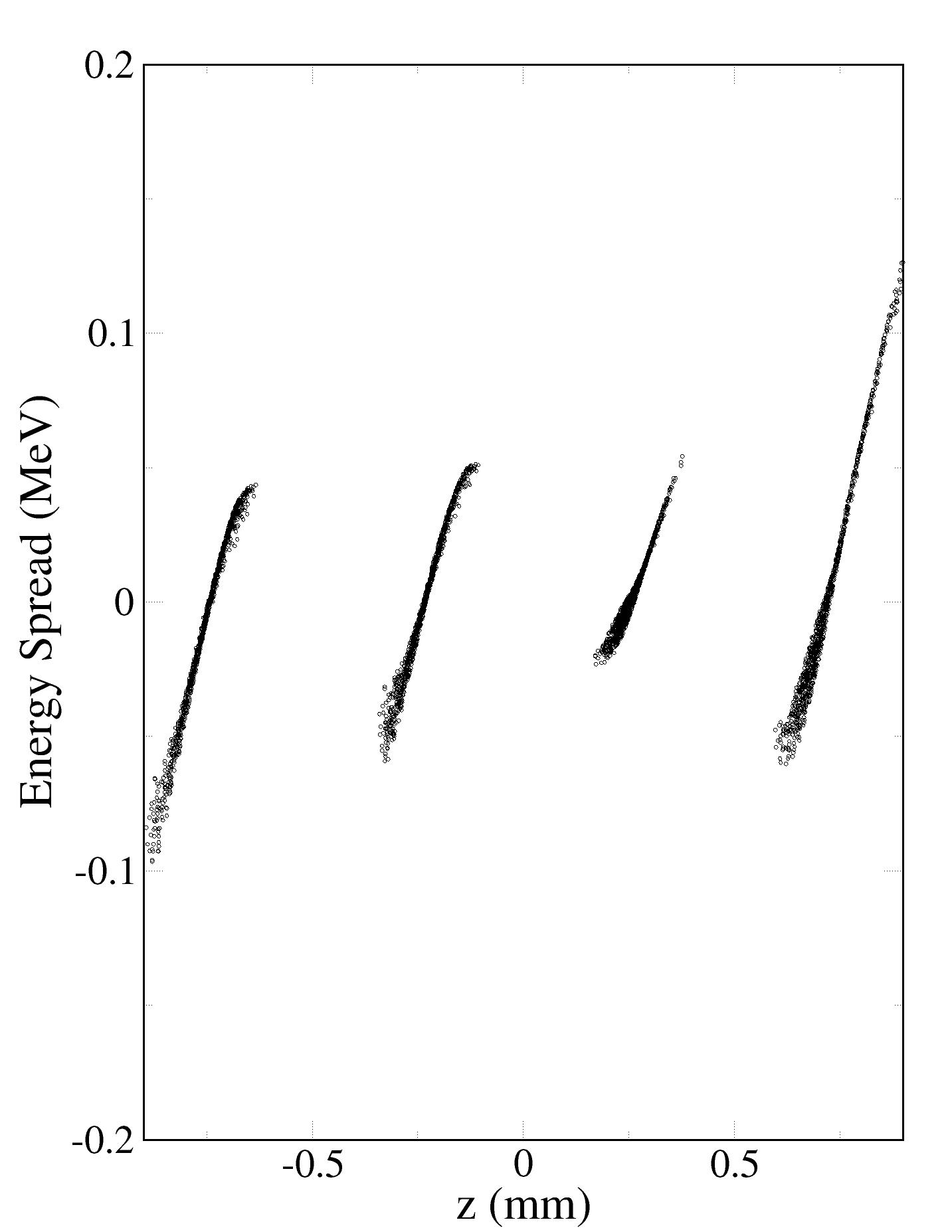}\includegraphics[height=5.7cm,keepaspectratio]{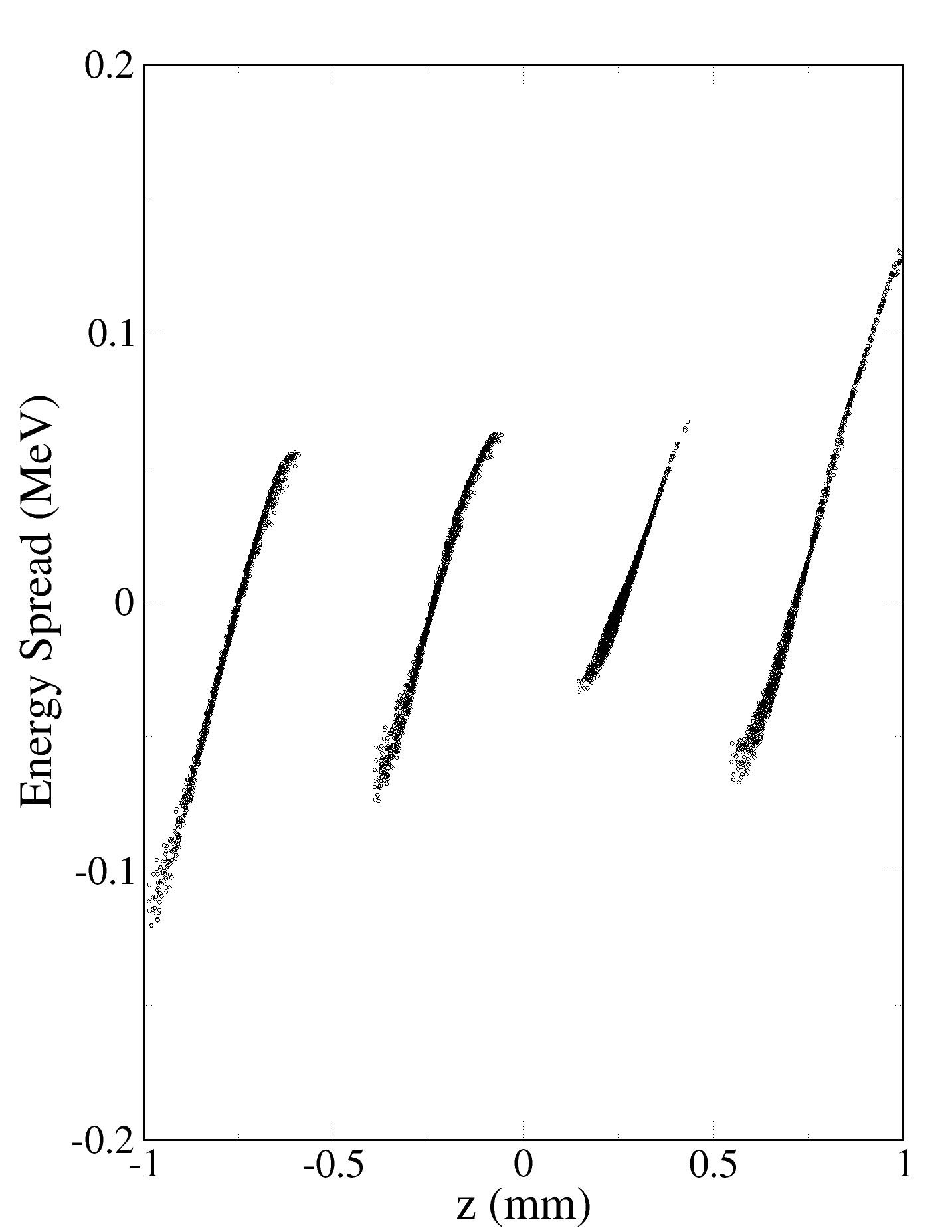}\\
\includegraphics[height=5.7cm,keepaspectratio]{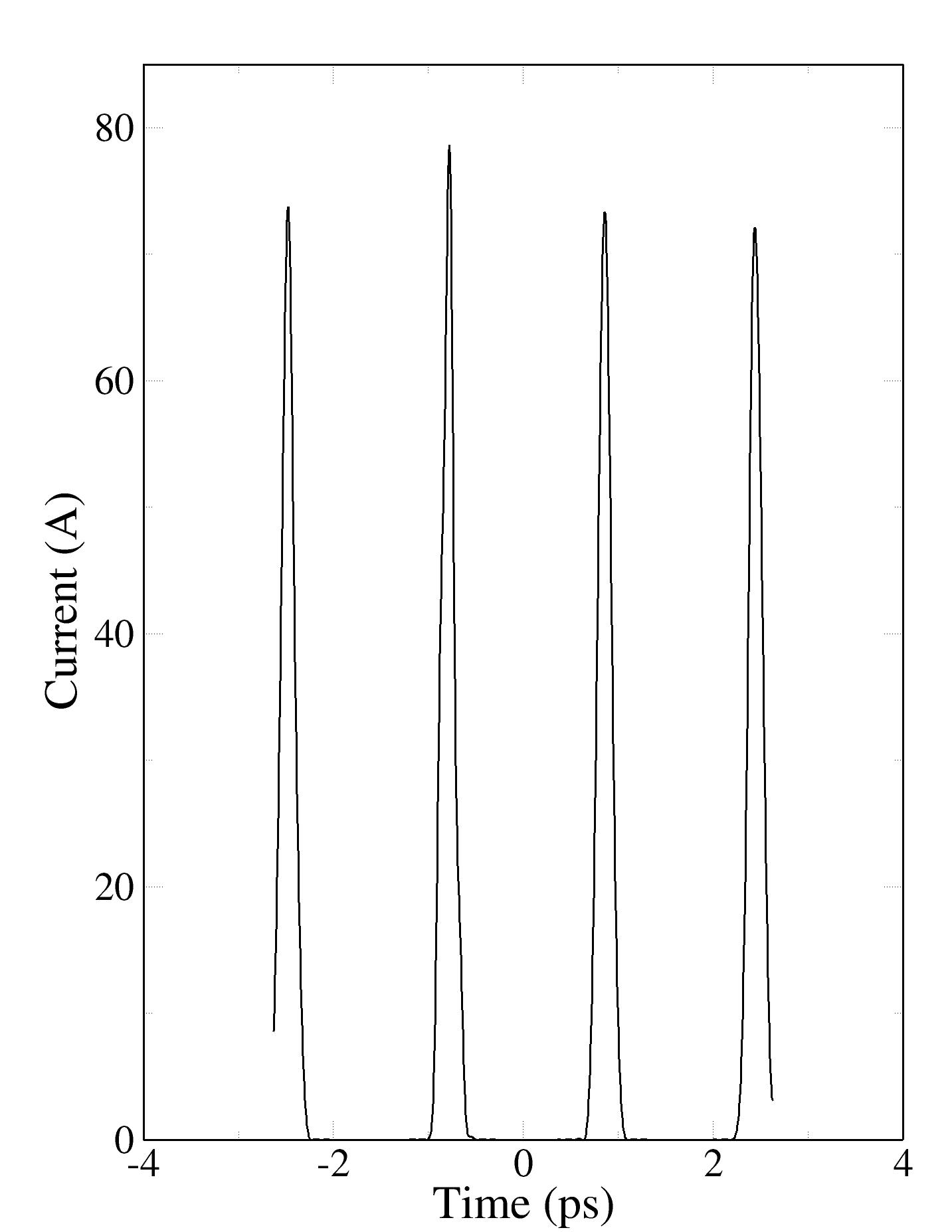}\includegraphics[height=5.7cm,keepaspectratio]{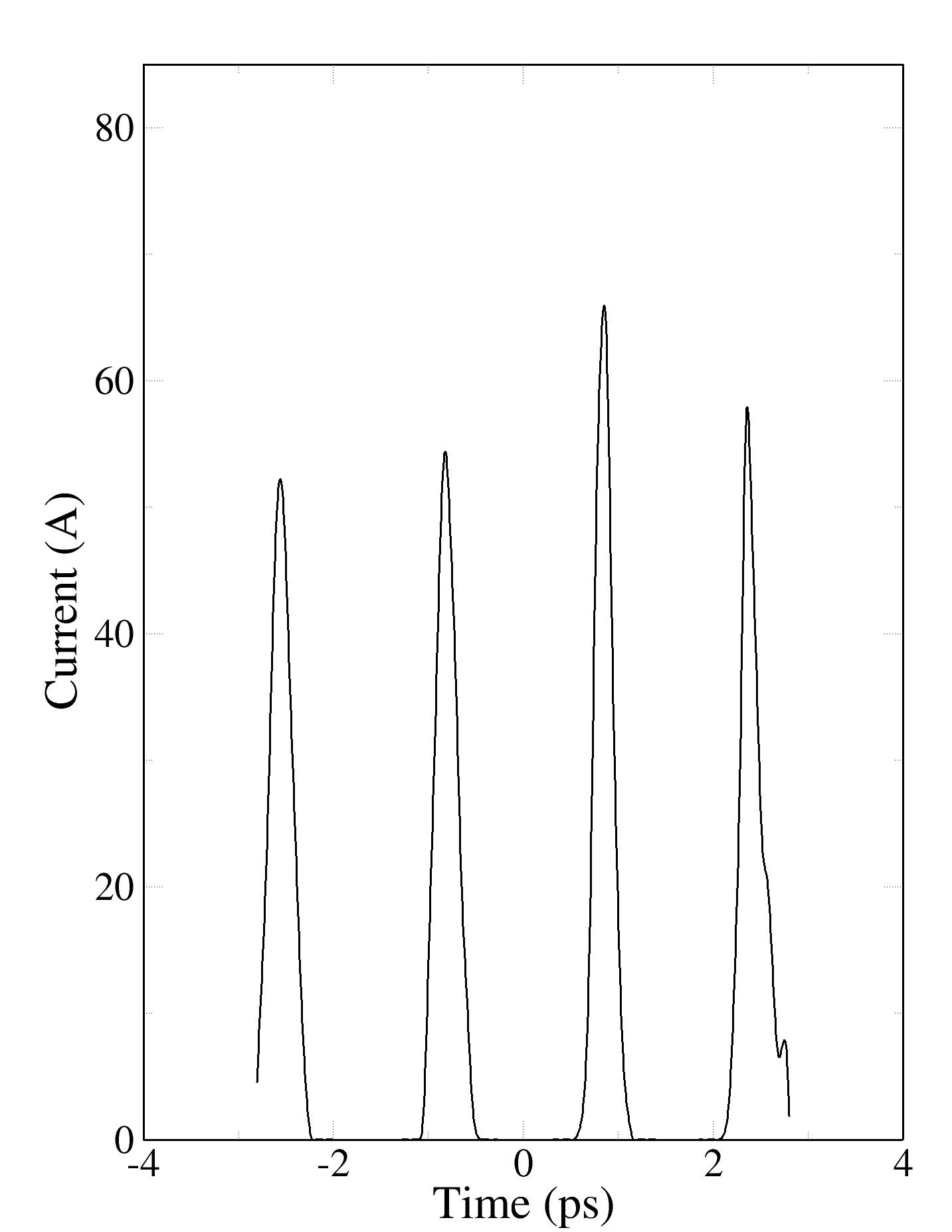}\includegraphics[height=5.7cm,keepaspectratio]{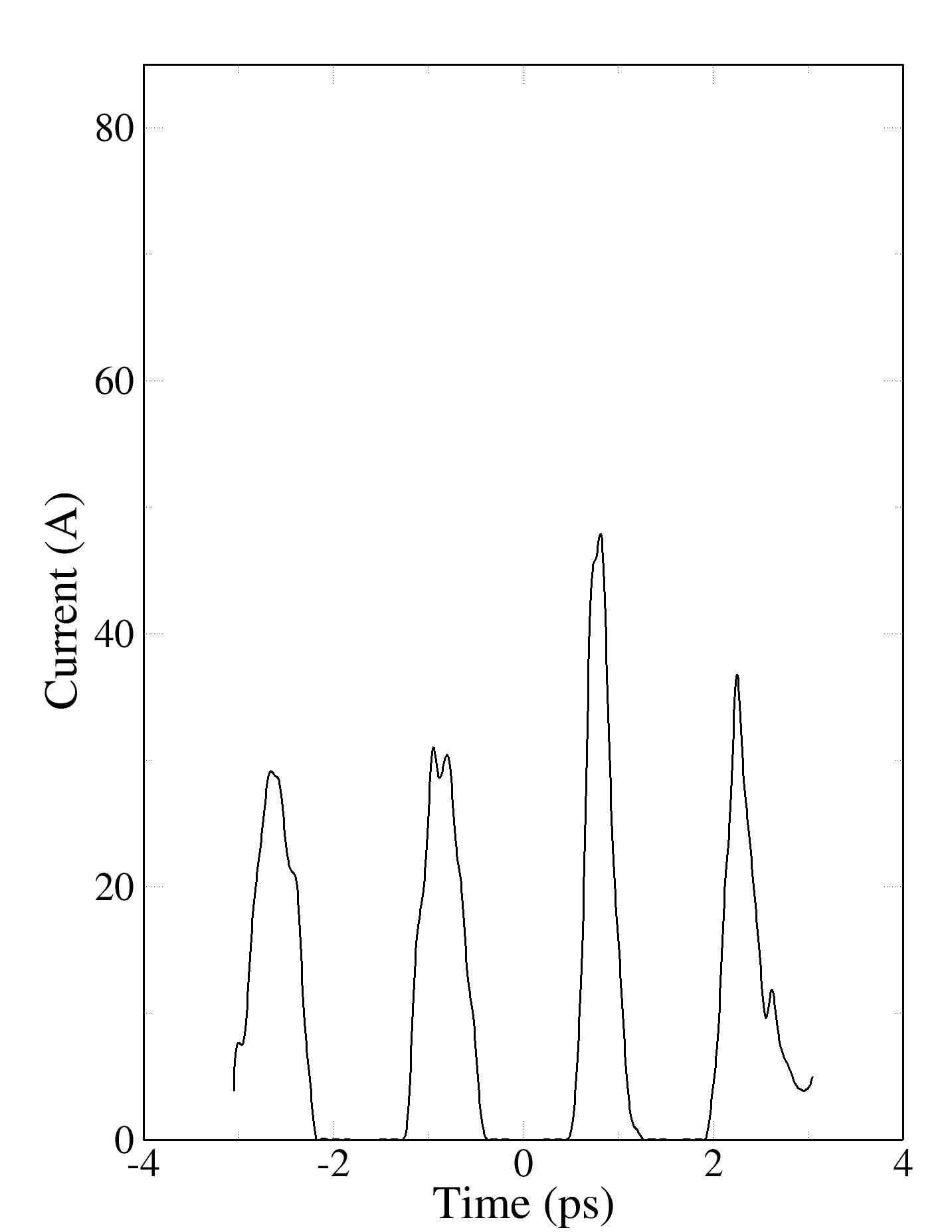}\includegraphics[height=5.7cm,keepaspectratio]{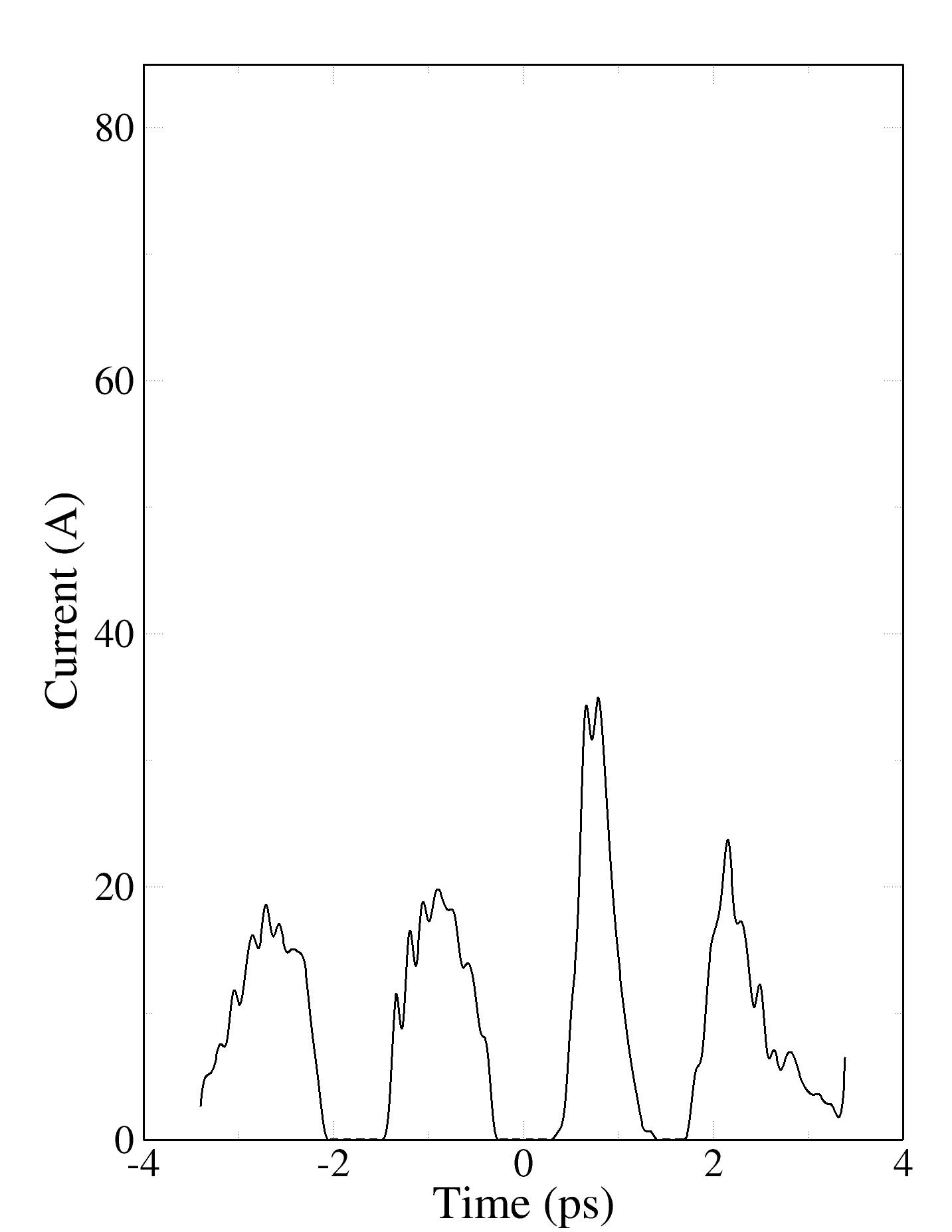}
\caption{\label{fig:06_beamoptics}Snapshot of electron beam profile (Top), Energy Spread (Middle) and Peak Current Distribution (Bottom) for 0.6 THz at (a) Photocathode (b) Undulator's entrance (c) Undulator's centre and (d) Undulator's exit.}
\includegraphics[width=5.43cm,keepaspectratio]{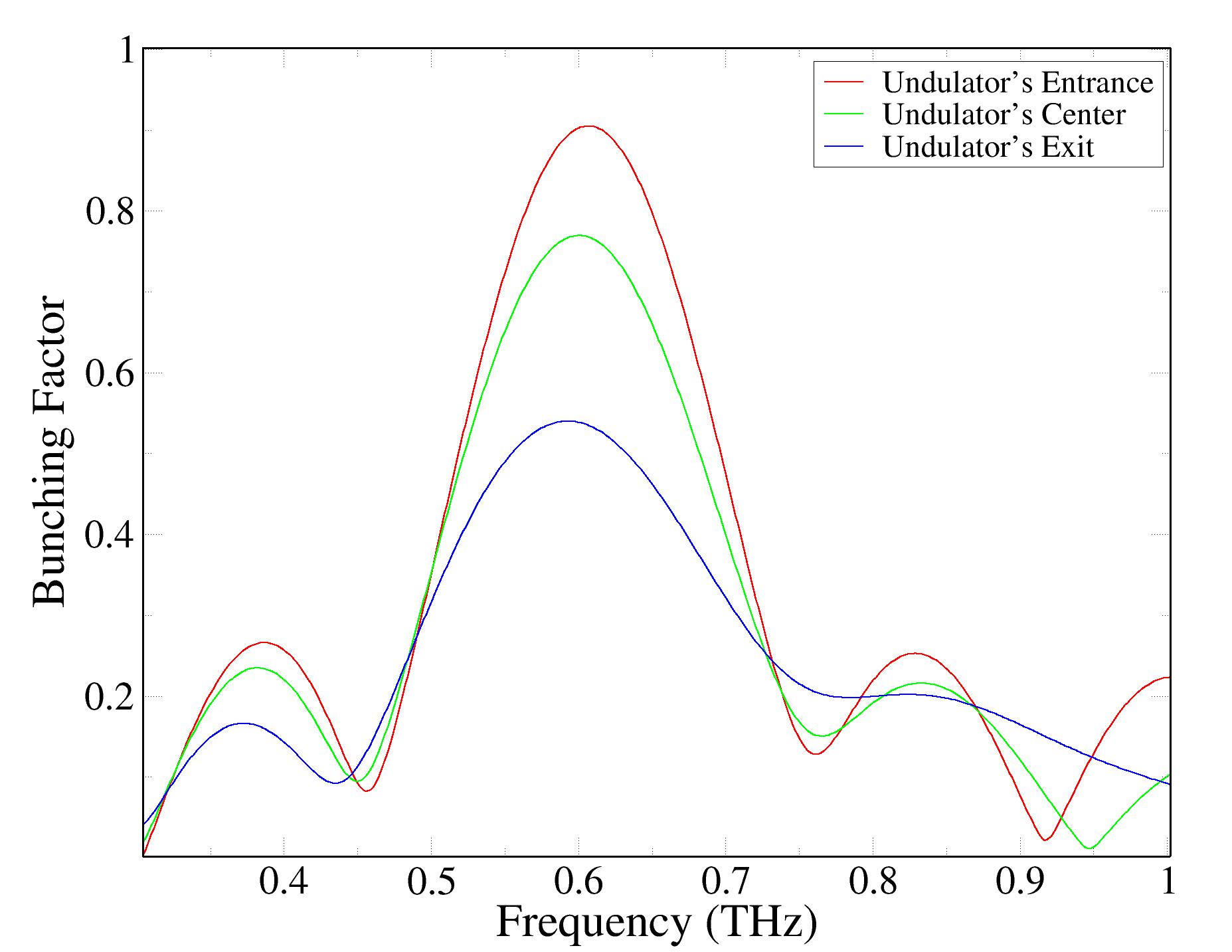}\includegraphics[width=5.43cm,keepaspectratio]{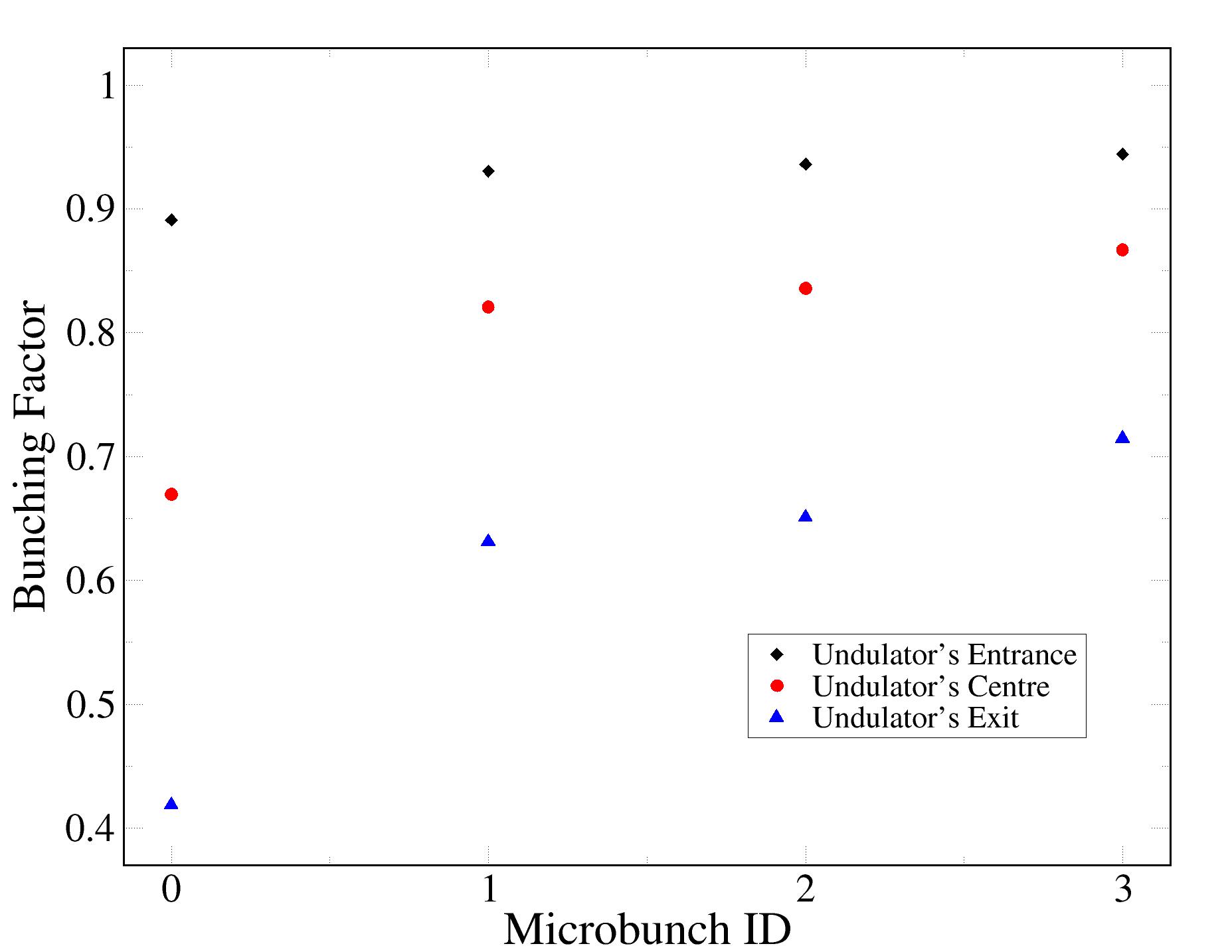}\includegraphics[width=5.43cm,keepaspectratio]{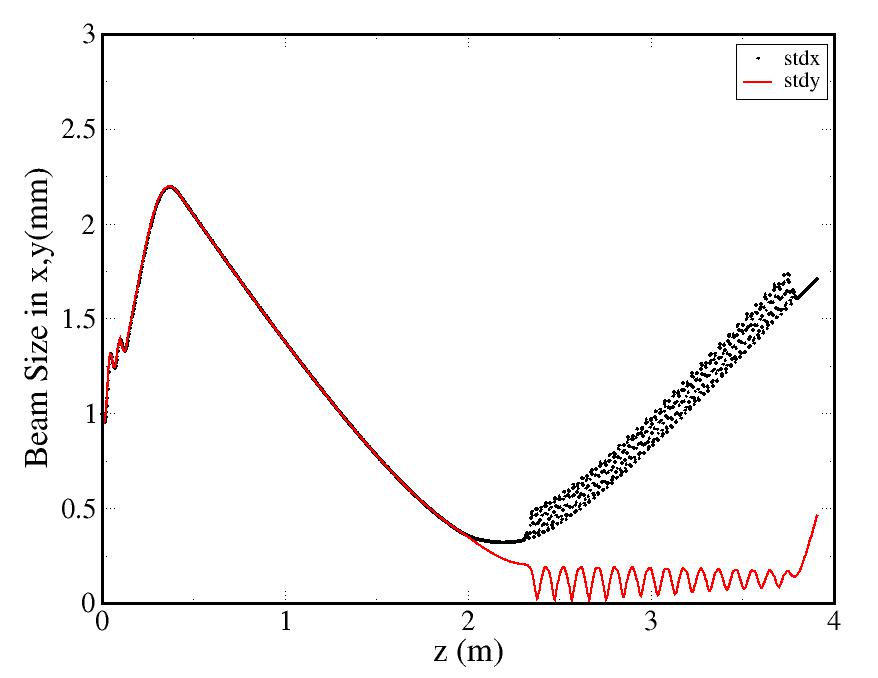}
\caption{\label{fig:dls_06THZ_bf}Variation of the comb beam average bunching factor (left), individual bunching factor (middle) and envelope through the beamline for 0.6 THz (right).}
\end{figure*}
For the case of 0.6 THz, only four micro-bunches separated by 2.2 ps (at the photocathode) can be produced (shown in Fig. \ref{fig:06_beamoptics}). The growth of energy spread of individual micro-bunches from the photocathode to the undulator's entrance is almost $\sim$50 keV. It is seen that during the transportation of the comb beam from photocathode to the undulator over a distance of 2.3 m, the changes in the temporal structure of the comb beam due to space charge forces is negligible. However; inside the undulator, the micro-bunches experience dispersion (large K $\sim$2.48) and tend to lose their longitudinal bunching. This is reflected in the drop of bunching factor from unity at the undulator's entrance to about 0.5 at the undulator's exit over a distance of 1.5m only (Fig. \ref{fig:dls_06THZ_bf}(left)). Thus, the peak current drops from $\sim$60 Amperes at the undulator's entrance to $\sim$20 Amperes at the undulator's exit. The evolution of the electron beam envelope is shown in Fig. \ref{fig:dls_06THZ_bf} (right).

\section{THz Radiation Simulations} \label{THz calculations}
To evaluate the various parameters of THz radiation emitted from the wiggling electrons inside the undulator, a simulation code written in $C^{++}$ has been developed \cite{Lehnert2017}. The code solves for the radiation wave packet emitted from the ensemble of the charged particles moving under the influence of arbitrary electromagnetic fields by solving the Li\~enard-Wiechert fields for point particles (Eq. \ref{LW}). In the code, the particles are loaded and tracked when they are moving through the undulator's field using Vay Algorithm. The algorithm differs from the Boris Pusher (which is the default particle pusher used for integrating equations of motion of charged particle in electromagnetic fields) in the manner it averages velocity (from $'$old$'$ and $'$new$'$ velocities) to give magnetic rotation to the particle \cite{Vay2008,Boris1971,Higuera2017}. The entire history of the trajectory and information like velocity, momentum, acceleration and position of the particles inside the undulator are recorded into the SDDS output files \cite{Borland1995}. In order to calculate the fields, firstly the retarded time at which the radiation was emitted is calculated and then the relative distance between the source of radiation and observation point, velocity and acceleration of the particle are evaluated at that retarded time.\par
The undulator used for simulation is a 30 period planar geometry and has a total length of 1.5 m. The magnetic field due to end pieces of the undulator used for simulations has been modelled with a linear ramp (extending over an undulator period) superimposed on the undulator$'$s sinusoidal field. This ensures zero displacement of a particle injected on the axis of the undulator. In present simulations, the interaction between the electrons and between the electron-radiation have been ignored. The important parameters used for THz simulations are shown in Table \ref{tab:table2}.\par
\begin{figure}[h]
	\includegraphics[width=8.25cm]{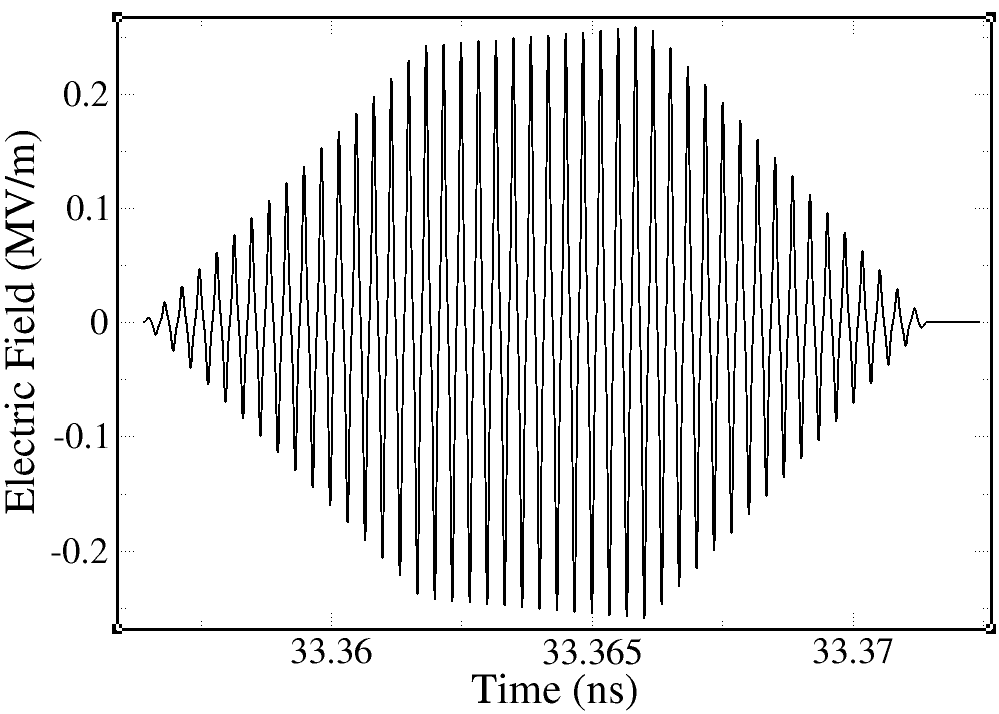}
	\includegraphics[width=8.25cm]{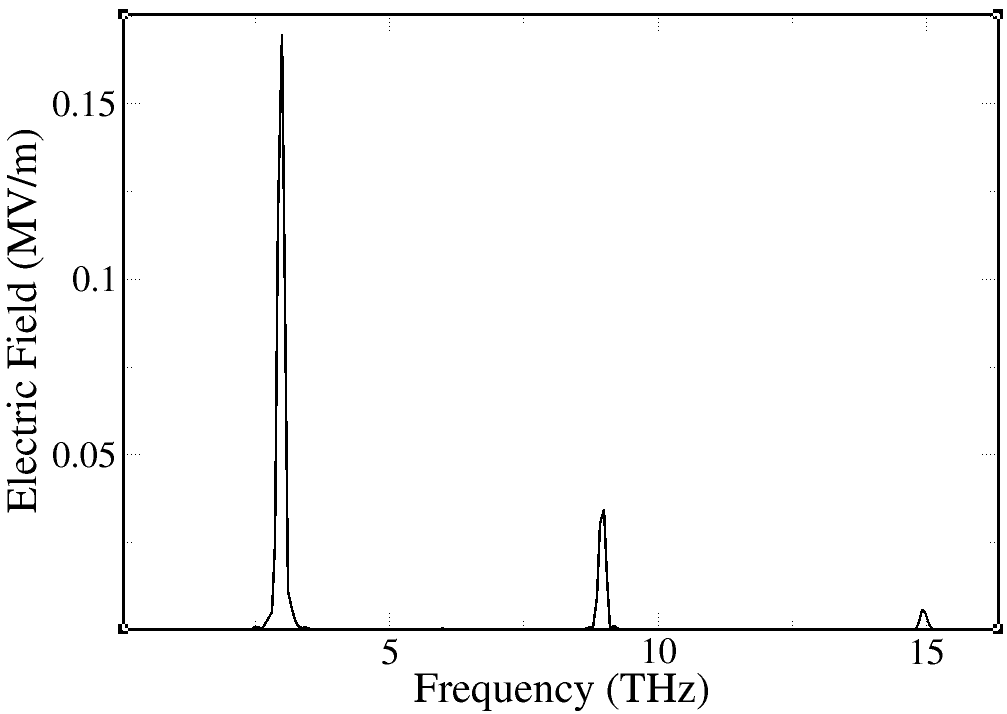}
	\caption{\label{fig:16micro_ideal}Radiation Wave packet emitted from an ideal comb beam having a train of 16 micro-bunches (Top) and its spectrum (Bottom) evaluated at 10m from the undulator's entrance.}
\end{figure}

\begin{table}[h]
	\caption{\label{tab:table2}%
		Parameters Used for THz Calculations
	}
	\begin{ruledtabular}
		\begin{tabular}{cccc}
			\textrm{Freq.(THz)}&
			\textrm{Electron Energy (Mev)}&
			\textrm{K}&
			\textrm{$B_u (T)$}\\
			\colrule
			0.6&6.65&2.48&0.52\\
			3 &8.1&	0.6&	0.128\\
		\end{tabular}
	\end{ruledtabular}
\end{table}
It was earlier explained that the maximum output radiation from the pre-bunched electron beam is obtained when each micro-bunch is super-radiant i.e. B($\omega$)$\rightarrow$1 and the separation is exactly equal to the wavelength of the fundamental undulator radiation. If both the conditions are satisfied then the electron beam can be said to be an ideal comb beam. An ideal comb beam (corresponding to 3 THz) with 16 micro-bunches and 15 pC/micro-bunch charge was generated and injected into the undulator. The separation was kept exactly equal to 0.1 mm i.e. the wavelength corresponding to 3 THz radiation. The output radiation waveform was evaluated at 10m from the undulator's exit and is shown in Fig. \ref{fig:16micro_ideal}(Top). It can be seen that the growth in the field amplitude is exactly linear and the number of periods required to achieve saturation is exactly equal to the number of micro-bunches (Fig. \ref{fig:dls_schematic}). Further, the duration of the pulse can also be shown to be close to $T_{pulse}  =\lambda_r (N_u+N_m/\beta)/c$, which is equal to 15.34 ps. Fig. \ref{fig:16micro_ideal} (bottom) shows the spectrum of the radiation obtained from the Fourier Transform of the radiation wave packet and it is seen that the first and the third harmonic frequency content is present, while even harmonics are absent. This can be understood from the longitudinal and the transverse motion of the electron inside the undulator \cite{Schmuser2008,Clarke2004}. For high frequencies, the undulator will be tuned to K \textless 1 where the electron's deflection from its path is low and the device acts as an undulator. For the longer wavelengths, however, the K parameter will be set up to $\sim$2.8 where the device can be described as a wiggler. For a point detector, the entire radiated wave packet will be detected only if the deflection angle of the electron beam is small compared to $1/\gamma$. Then, the electric field of the emitted radiation wavepacket will follow a pure sinusoidal curve; whose Fourier Transform will have only a single frequency component; resulting only in the fundamental harmonic of the radiation. However, as K increases, the transverse displacement of the electron beam increases and the point detector is unable to detect the complete radiation waveform. It only sees a series of short field pulses periodic in time which are emitted at the extremes of the trajectory when the velocity vector points towards the observer. The polarity depends on the sign of the transverse acceleration of the electrons. This behaviour is due to the $1/(1-\boldsymbol{\hat{n}}\cdot \boldsymbol{\beta})^3$ term of Eq. \ref{LW} which becomes very large only if $\boldsymbol\hat{n}$ $\cdot$ $\boldsymbol{\beta}=1$. The detected radiation waveform is then described by a function which is symmetric and uniformly spaced in time i.e. it can be defined by a function like f(t+T)=f(t). Further, the obtained waveform is such that it reverses its polarity after every half-period i.e. f(t+T/2)=f(t). Thus, the function describing the radiation wave packet is such that its Fourier transform will have odd-harmonics only. If the observation is extended off the undulator axis the even harmonics reappear. \par
\begin{figure}[h]
	\includegraphics[width=8.25cm]{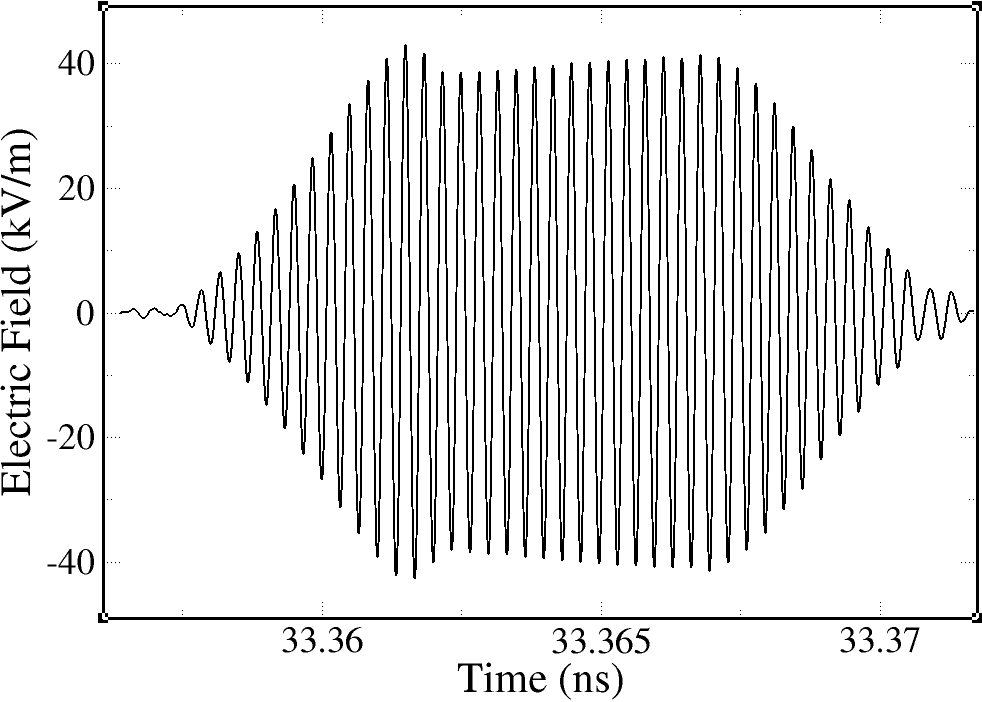}
	\includegraphics[width=8.25cm]{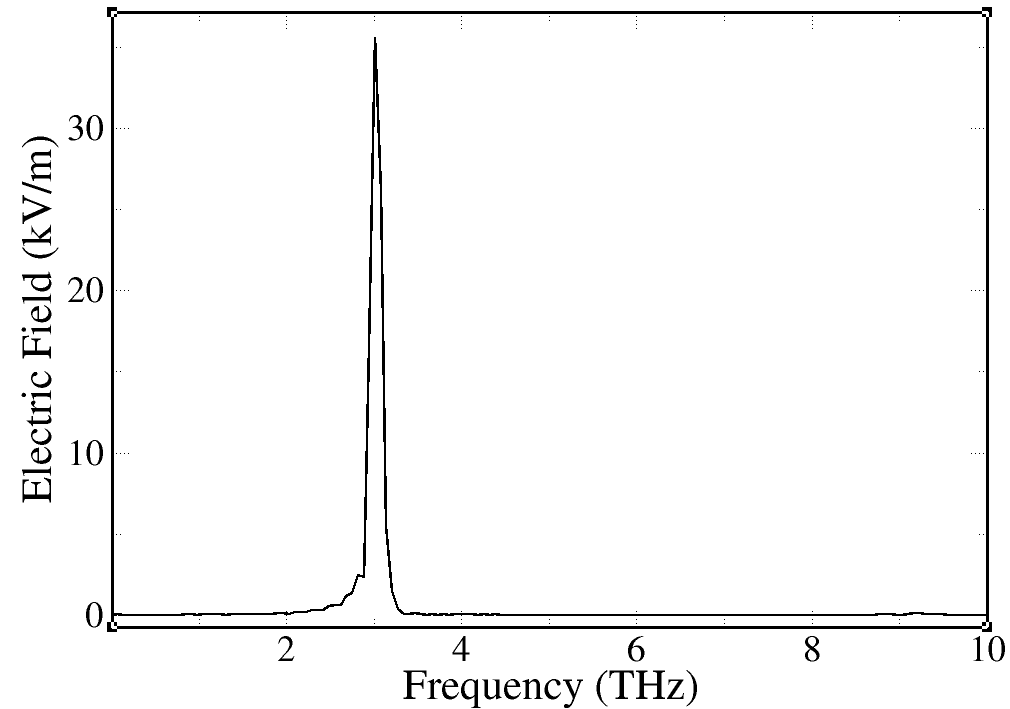}
	\caption{\label{fig:16micro_real}Radiation Wave packet emitted from a comb beam \big(with $B_{avg}(\omega =$ 3 THz) $\sim$ 0.2\big) having a train of 16 micro-bunches (Top) and its spectrum (Bottom) evaluated at 10m from the undulator's entrance.}
\end{figure}

The same undulator configuration with an observer placed at 10 m, was also used to study the radiation emitted from the micro-bunched structure as shown in Fig. \ref{fig:3THz_beamoptics}. The 6-D phase space information (x, px, y, py, z, pz) of the electron beam at the undulator's entrance was extracted from the GPT code and imported into the $C^{++}$ code. The radiation output obtained from the actual 16 micro-bunch structure is shown in Fig. \ref{fig:16micro_real}(top). In this case, it is observed that the growth of radiation in first few undulator periods is slow and non-linear. This is because the first micro-bunch that enters the undulator is completely de-bunched (as evident from its individual bunching factor in Fig. \ref{fig:dls_3THZ_bf}).
The radiation emitted from the first micro-bunch is essentially incoherent and is equivalent to the $'$shot noise$'$ of the FEL. As soon as the super-radiant micro-bunches enter the undulator, they emit coherent radiation and linear growth in the field is observed again. \par
\begin{figure}[h]
	\includegraphics[width=8.25cm]{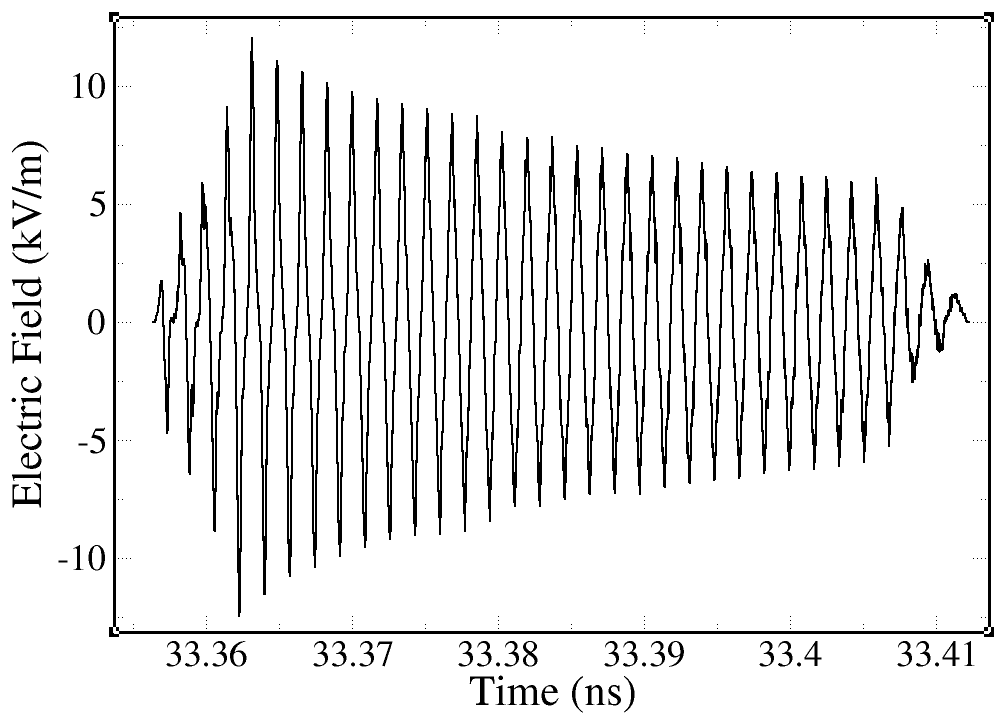}
	\includegraphics[width=8.25cm]{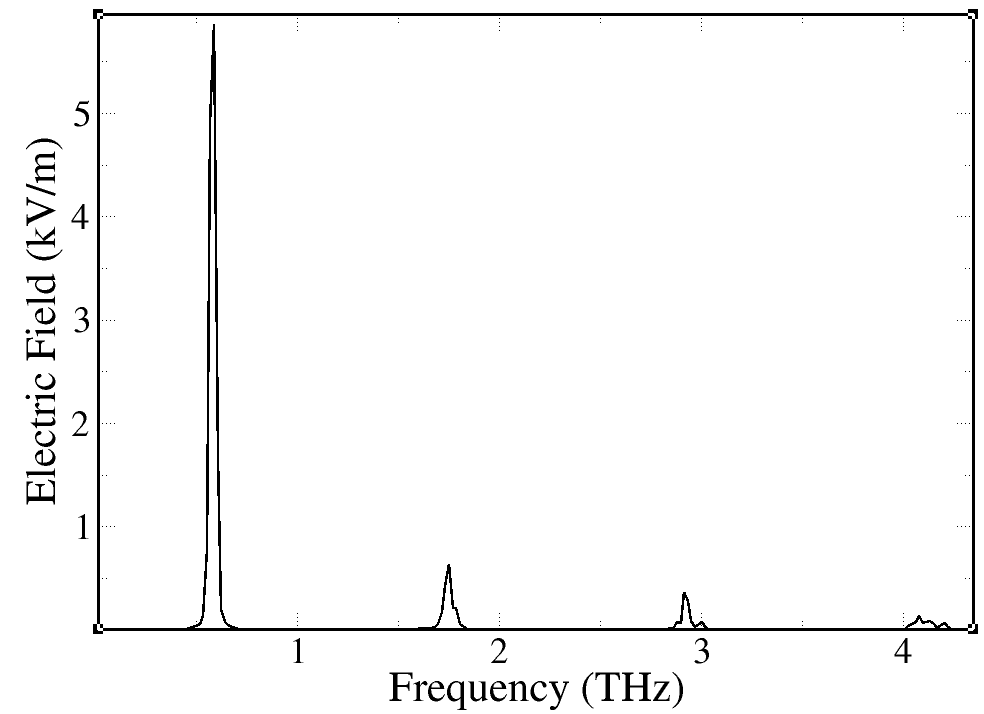}
	\caption{\label{fig:4micro_real}Radiation Wave packet emitted from a comb beam (with $B_{avg}(\omega =$ 0.6 THz) $\sim$ 0.77) having a train of 4 micro-bunches (Top) and its spectrum (Bottom) evaluated at 10m from the undulator's entrance.}
\end{figure}

For the case of 0.6 THz, the radiation output from electron beam (Fig. \ref{fig:06_beamoptics}) is shown in Fig. \ref{fig:4micro_real}. From Fig. \ref{fig:06_beamoptics}, it can be seen that the drop in the bunching factor and current profile is drastic for this case. Thus, the electric field amplitude tend to decrease (rather than saturating) in output radiation from this case. In this case, the undulator parameter is large i.e. K $\sim$ 2.43 and therefore, the spectrum obtained has some contribution from the higher harmonics as well. The spectral power density for the frequency of interest is low.\par
\begin{figure}[h]
	\includegraphics[height=7.75cm,width=8.25cm]{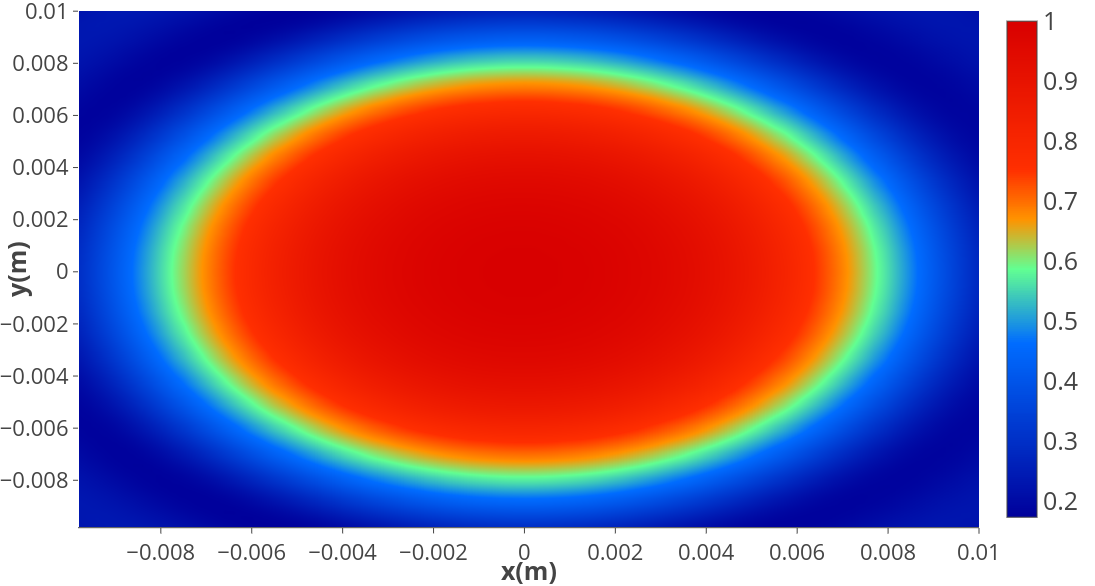}
	\includegraphics[height=7.75cm,width=8.25cm]{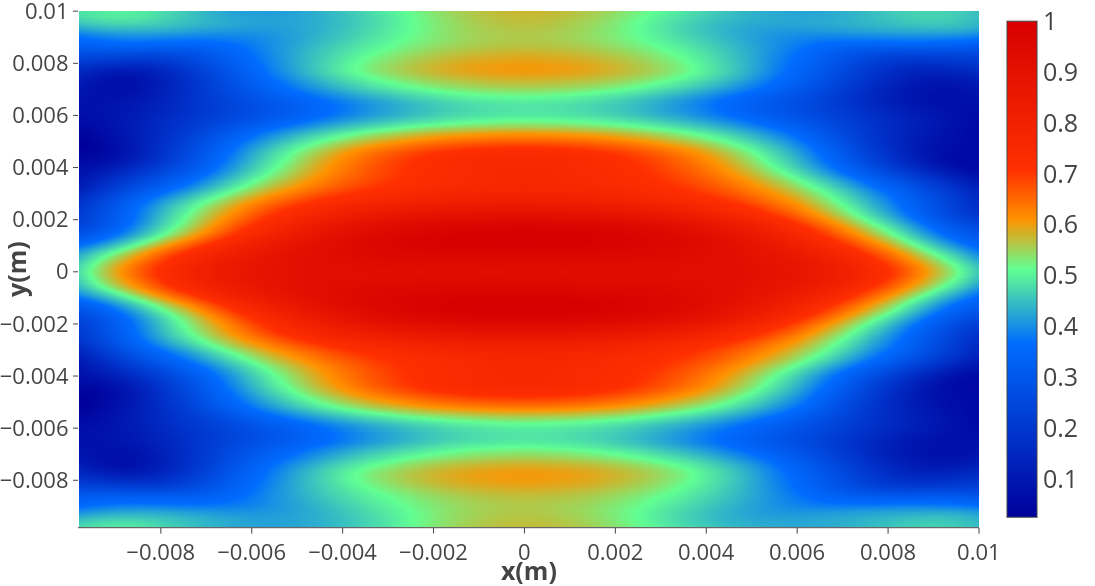}
	\captionof{figure}{\label{fig:16micro_profile} Normalized Radiation Intensity obtained (assuming free space propagation) for 3 THz on a 20 $\times$ 20 mm$^2$ grid placed at 0.5 m from undulator's exit (top) and distortion of the transverse radiation beam profile due to multiple reflections on the wall (bottom).}
\end{figure}

 The transverse radiation beam profile of the THz radiation from pre-bunched electron beam having energies of the order of few MeVs is an important point to consider. The radiation beam profile is evaluated at 0.5 m away from the undulator's exit (Fig. \ref{fig:16micro_profile}-top) and this is the place where the beam pipe of dimension 40 mm (x) $\times$ 20 mm (y) placed inside the undulator is connected to another beam pipe of much larger dimension.  It is understood that the radiation emitted by the electron beam under large angles hits the vacuum chamber walls and undergoes multiple reflections. The reflected waves will interfere with the core beam to produce the radiation pattern at the output of the device.
To simulate this effect, we introduce mirror particles to all the tracked particles, which are displaced by multiples of the vacuum chamber height and have alternating polarities owing to the 180 degree phase jump; a wave undergoes upon reflection at a metallic surface. The radiation of the mirror particles is delayed due to their larger distance to the observer. The radiation of all ($'$real$'$ and $'$mirror$'$) particles is added coherently to obtain the total radiation output. This description is valid only inside the waveguide aperture so we have to compute the radiation pattern at the end of the narrow undulator vacuum chamber and use another (free-space) optical propagation code to simulate the radiation propagation from there.\par
At 3 THz, the central cone of the radiation just fits into the opening angle of the vacuum chamber aperture, so the wave guiding effect is small as can be seen from Fig. \ref{fig:16micro_profile}(bottom). The radiation of 0.6 THz corresponds to a longer wavelength (produced using electrons of lower energy) and therefore; larger divergence angle yields a much more pronounced interference pattern. It is best described as the sum of the few lowest-order transverse waveguide modes \cite{Elias1983}. One should note that the phase velocity of the waveguide modes slightly differs from the vacuum speed of light so the resonance frequency of the undulator radiation shifts accordingly.
\\The radiation pulse energy emitted into the central cone of the undulator is strongly dependent on the deflection angle of the electron beam through undulator parameter K, the strength of correlation among the radiating charges through the bunching factor $B_{avg}(\omega$) and the total number of electrons through the total charge of the bunch Q. It can be computed using the following equation \cite{Bocek1996}:
\begin{equation}
E_{cen}=B_{avg}^2 (\omega)\dfrac{\pi Q^2}{2\epsilon_0\lambda}\dfrac{K^2}{1+K^2/2}JJ(K)
\end{equation}

where $JJ(K)=(J_0 (\xi)- J_1 (\xi))^2$  is called the coupling factor of the electron beam to the radiation field and $\xi=K^2/(4+K^2/2)$.  Table \ref{tab:table3}, lists the expected optical pulse energies emitted into the central cone by the comb beam obtained for various comb beam configurations and has been compared to ideal comb beam. It is to be noted that although the higher frequencies have much higher total charge in the comb beam; but; as the bunching factor is lower for these frequencies, the total pulse energy is remarkably low.

\begin{table}[h]
	\caption{\label{tab:table3}%
		Optical pulse energy emitted into the central cone of the undulator radiation by an ideal and a realistic comb electron beam
	}
	\begin{ruledtabular}
		\begin{tabular}{cccccc}
			\multicolumn{1}{p{1cm}}{\centering Frequency \\ $\left(THz\right)$}&
			\multicolumn{1}{p{1cm}}{\centering K}&
			\multicolumn{1}{p{1.5cm}}{\centering $Q_{total} (pC)$, \\ $N_m$}&
			\multicolumn{1}{p{1cm}}{\centering $B_{avg}(\omega)$}&
			\multicolumn{1}{p{1cm}}{\centering $E_{cen}^{Ideal}$\\ ($\mu$J)}&
			\multicolumn{1}{p{1cm}}{\centering $E_{cen}^{Real}$\\ ($\mu$J)}\\
			\colrule
			0.18&	3.0&	30, 2&	0.45&	0.08&	0.044\\
			0.6	&2.43&	60, 4&	0.77&	1.17&	0.7\\
			1.00&	1.2&	120, 8&	0.72&	4.70&	2.88\\
			1.25&	1.05&	120, 8&	0.55&	6.18&	1.85\\
			1.50&	1.0&	120, 8&	0.52&	3.37&	1.88\\
			2.00&	0.99&	120, 8&	0.43&	9.32&	1.75\\
			2.50&	0.81&	240, 16&	0.37&	37.0&	5.03\\
			3.00&	0.6&	240, 16&	0.20&	19.1 &	1.15
			
		\end{tabular}
	\end{ruledtabular}
\end{table}
\section{Conclusion}
In the paper, the feasibility study of generating the comb electron beam by manipulating the laser pulses incident on the photocathode is reported. We have shown that the bunching factor of a comb beam will depend on the structure of the individual micro-bunches and also on the separation between the centre of masses of the micro-bunches. The simulation calculation shows that the average bunching factor of the 16 micro-bunch train can be maintained at a value of $\sim$0.2 from the entrance to the exit of the undulator. Except for the incoherent radiation emitted from the first micro-bunch, the output radiation waveform for 3  THz shows that the radiation wave packet emitted from trailing micro-bunches interfere constructively with the radiation wave packets emitted by leading micro-bunches and therefore radiation field grows linearly. For larger wavelengths, the number of micro-bunches that can be accomodated in a single micro-bunch train is reduced. The number of micro-bunches for 0.6 THz is only 4. The energy of the comb beam required to generate longer wavelengths is lesser, the undulator parameter is large and the ratio of energy spread to that of average energy of the beam i.e. $\Delta$E/E is sufficiently large to strongly disperse the microbunches inside the undulator.  The effect of the dispersion is that the electrons become uncorrelated and the radiation enhancement due to $"$super-radiance$"$ is lost to some extent. The study of the transverse beam profile of the THz radiation and the effect of beam-pipe inside the undulator on the beam profile suggests that there will be some distortion due to the beam pipe; suggesting the requirement of cylindrical lenses to focus the beam at desired location. For longer wavelengths, the beam-pipe will act like a wave-guide and therefore a wave-guide analysis is required. At last, the radiation pulse energy emitted into the central cone of undulator has been calculated for the entire range of the frequencies.

\section{Acknowledgment}
One of the authors, S. Tripathi (PH/16-17/0029) would like to acknowledge University Grant Commission (UGC), New Delhi, India for financial support as D.S.Kothari Postdoctoral fellowship.
The authors of IUAC would like to acknowledge Board of Research in Nuclear Sciences (BRNS), India for providing financial support and Prof. Peter Michel, HZDR for arranging the visit of Mr. Vipul Joshi to ELBE facility. The help received from  Mr. Rohan Biswas, IUAC for fruitful discussions on the theory of emission of radiation from the comb beam is appreciated.

\bibliography{library}

\end{document}